\documentclass[twocolumn,pra,aps,10pt,superscriptaddress]{revtex4-1}

\usepackage{graphicx}
\usepackage{amsmath}
\usepackage{amsfonts}
\usepackage{amssymb}
\usepackage{xcolor}
\usepackage{hyperref}
\hypersetup{colorlinks=true,allcolors=blue}
\usepackage[T1]{fontenc}

\usepackage{footnote}
\usepackage{textcomp}

\newcommand{\mbf}[1]{\mathbf{#1}}
\renewcommand{\t}[1]{\textrm{#1}}
\newcommand{\nn}{\nonumber\\}

\newcommand{\q}{\mbf{q}}

\newcommand{\g}{\gamma}

\renewcommand{\r}{\rho}
\newcommand{\s}{\sigma}

\newcommand{\w}{\omega}

\newcommand{\D}{\Delta}

\renewcommand{\L}{\mathcal{L}}

\newcommand{\+}{^\dagger}
\renewcommand{\>}{\rangle}
\newcommand{\<}{\langle}
\newcommand{\Tr}{\t{Tr}}

\newcommand{\G}{\vert G\>}
\newcommand{\X}{\vert X\>}
\newcommand{\XL}{\vert X_L\>}
\newcommand{\XR}{\vert X_R\>}
\newcommand{\B}{\vert B\>}
\newcommand{\XI}{\vert X_1\>}

\newcommand{\XX}{\vert X\>\<X\vert}
\newcommand{\XLXL}{\vert X_L\>\<X_L\vert}
\newcommand{\XRXR}{\vert X_R\>\<X_R\vert}
\newcommand{\XIXI}{\vert X_1\>\<X_1\vert}
\newcommand{\BB}{\vert B\>\<B\vert}
\newcommand{\XHXH}{\vert X_H\>\<X_H\vert}

\newcommand{\GX}{\vert G\>\<X\vert}

\newcommand{\GXL}{\vert G\>\<X_L\vert}

\newcommand{\GXR}{\vert G\>\<X_R\vert}

\newcommand{\GXI}{\vert G\>\<X_1\vert}

\newcommand{\GXH}{\vert G\>\<X_H\vert}

\newcommand{\XLB}{\vert X_L\>\<B\vert}

\newcommand{\XRB}{\vert X_R\>\<B\vert}

\newcommand{\XHB}{\vert X_H\>\<B\vert}

\newcommand{\XLXR}{\vert X_L\>\<X_R\vert}
\newcommand{\XRXL}{\vert X_R\>\<X_L\vert}

\begin{document}

\title{On-demand generation of higher-order Fock states in quantum-dot--cavity systems}


\author{M. Cosacchi}
\affiliation{Theoretische Physik III, Universit{\"a}t Bayreuth, 95440 Bayreuth, Germany}
\author{J. Wiercinski}
\affiliation{Theoretische Physik III, Universit{\"a}t Bayreuth, 95440 Bayreuth, Germany}
\author{T. Seidelmann}
\affiliation{Theoretische Physik III, Universit{\"a}t Bayreuth, 95440 Bayreuth, Germany}
\author{M. Cygorek}
\affiliation{Heriot-Watt University, Edinburgh EH14 4AS, United Kingdom}
\author{A. Vagov}
\affiliation{Theoretische Physik III, Universit{\"a}t Bayreuth, 95440 Bayreuth, Germany}
\affiliation{ITMO University, St. Petersburg, 197101, Russia}
\author{D. E. Reiter}
\affiliation{Institut f\"ur Festk\"orpertheorie, Universit\"at M\"unster, 48149 M\"unster, Germany}
\author{V. M. Axt}
\affiliation{Theoretische Physik III, Universit{\"a}t Bayreuth, 95440 Bayreuth, Germany}

\begin{abstract}
The on-demand preparation of higher-order Fock states is of fundamental importance in quantum information sciences. We propose and compare different protocols to generate higher-order Fock states in solid state quantum-dot--cavity systems. The
protocols make use of a series of laser pulses to excite the quantum
dot exciton and off-resonant pulses to control the detuning between dot
and cavity.
Our theoretical studies include dot and cavity loss processes as well as the pure-dephasing type coupling to longitudinal acoustic phonons in a numerically complete fashion. By going beyond the two-level approximation for quantum dots, we  study the impact of a finite exchange splitting, the impact of a higher energetic exciton state, and an excitation with linearly polarized laser pulses leading to detrimental occupations of the biexciton state. We predict that under realistic conditions, a protocol which keeps the cavity at resonance with the quantum dot until the desired target state is reached is able to deliver fidelities to the Fock state $\vert 5\>$ well above $40\,\%$.

\end{abstract}

\maketitle

\section{Introduction}
\label{sec:Introduction}

Semiconductor quantum-dot--cavity (QDC) systems are widely discussed as candidates for highly integrable on-demand emitters of non-classical states of light.
They have been successfully proven to be reliable sources of high quality single photons \cite{Michler2000,Santori2001,Santori2002,
He2013,Wei2014Det,Ding2016,Somaschi2016,
Schweickert2018,Hanschke2018,Cosacchi2019}
as well as entangled photon pairs \cite{Akopian2006,Stevenson2006,Hafenbrak2007,
Dousse2010,delvalle2013dis,Mueller2014,Orieux2017,Seidelmann2019}.
Nonetheless, the preparation of higher-order Fock states remains a challenge.
These states find vast applications in quantum metrology  \cite{Holland1993,Nagata2007,Slussarenko2017}, as building blocks for more complex quantum states of light such as Schr\"odinger cat states \cite{Ourjoumtsev2007}, and in quantum computing \cite{Yamamoto1986}.

While schemes to prepare higher-order Fock states have been known in atomic cavity systems for decades \cite{Cummings1989,Varcoe2000,Haroche2012}, these protocols rely on properties specific to atoms, such as the finite time of flight through a resonator, which cannot be translated straightforwardly to a locally fixed solid state qubit as encountered in quantum dots (QDs).
Nonetheless, this protocol has been applied to a superconducting qubit coupled to a microwave cavity  \cite{Hofheinz2008}.
In this setup, coupling the qubit and the cavity only temporarily is achieved by changing the structure of the potential with an external flux bias and thus directly tuning the resonance frequency of the qubit. This is not possible in QDCs after the growth process is completed and thus the confinement potential set.
Furthermore, protocols involving parametric down-conversion have achieved remarkable fidelities to the targeted higher-order Fock states \cite{Tiedau2019}.
A huge challenge in such setups to be solved is the on-demand character of these sources.

In this work, we propose protocols for the preparation of higher-order Fock states in QDCs and explore their feasibility up to $n=5$. The protocols rely on the application of a series of ultrashort laser pulses combined with off-resonant laser pulses to induce an AC-Stark effect. In contrast to atoms, QDs are solid state systems and are therefore affected by the electron-phonon interaction. The pure-dephasing type coupling of the excitonic states to longitudinal acoustic phonons is known as the main source of decoherence in QDs even at cryogenic temperatures of a few Kelvin \cite{Besombes2001,Borri2001,Krummheuer2002,Axt2005,Reiter2019}. Accordingly, we study the influence of phonons as well as of cavity and radiative losses on the proposed protocols. Because we use ultrashort pulses, we further calculate the influence of higher energetic excitons on the preparation schemes.
In neutral QDs, the approximation of the QD as a two-level system is often reasonable.
In particular, this is the case when transitions to the biexciton are forbidden by selection rules.
Note that this sets constraints on the polarization of the driving laser as well as on the resonantly coupled cavity modes.
Furthermore, the two-level approximation holds well, when the fine-structure splitting (FSS) happens to be absent or is suppressed, e.g., by external fields, strain, or by fabricating highly symmetrical QDs \cite{Stevenson2006a,Young2006,
Stevenson2006b,Zhang2015,Cygorek2020}.
We study the respective influences by extending our system to a three- or four-level system. 

We show that even under these realistic conditions, our preparation schemes can reach fidelities to the Fock state $|5\>$ well above $40\,\%$. 
To put this value into perspective, let us compare it to other works.
Hofheinz \textit{et al.} \cite{Hofheinz2008} prepared Fock states with up to six photons in their superconducting qubit setup but do not report a value for the fidelity.
To be able to compare our results with theirs, we simulated the experiment described in Ref.~\onlinecite{Hofheinz2008} to reproduce the figures therein, while giving an estimate for the preparation fidelity of their setup.
For the fidelity to the state $\vert 5\>$ our estimate yields $\approx20\,\%$.
Tiedau \textit{et al.} \cite{Tiedau2019} use heralded parametric down-conversion to generate higher-order Fock states.
Their best fidelities to Fock states with $n\geq5$ do not exceed $50\,\%$.
Therefore, our QDC protocol is competitive with other means to prepare higher-order Fock states.
Compared with superconducting qubits, QDCs have the advantage of being in an energetic regime corresponding to the ps- rather than the ns-time scale, thus making the total preparation time for the Fock state $\vert 5\>$ about three orders of magnitude faster.

In the following, we start our analysis with a simple two-level QD-model and subsequently shift our focus to more complex situations by taking into account levels present in a QD that might have adverse effects on the preparation fidelity of higher-order Fock states.


\section{Protocols for a two-level system}
In atomic cavities, protocols for the preparation of higher-order Fock states have been successfully employed in the 1980s.
A well known example is the so called micromaser setup, where a highly excited Rydberg atom is brought to resonance with a single-mode cavity only during its time of flight through the resonator \cite{Cummings1989,Varcoe2000,Haroche2012}.
Sending excited atoms subsequently through the cavity fills the latter with one more photon at a time, thus preparing a higher-order Fock state.
While earlier experiments \cite{Varcoe2000} succeeded in the preparation of up to $n=2$, recent results yield states with up to seven photons \cite{Haroche2012}.
This technique was translated to a solid state platform by Hofheinz \textit{et al.} \cite{Hofheinz2008}.
In that work, a superconducting qubit is coupled to a single mode of a microwave cavity.
The transition frequency of the former can be tuned to bring it to resonance with the resonator frequency only during a finite time window, in which half a Rabi oscillation transfers the excitation from the qubit to a cavity photon.
This procedure simulates the finite dwell time of an atom in a cavity with a locally fixed superconducting qubit.

\begin{figure}[t]
	\centering
	\includegraphics[width=0.45\textwidth]{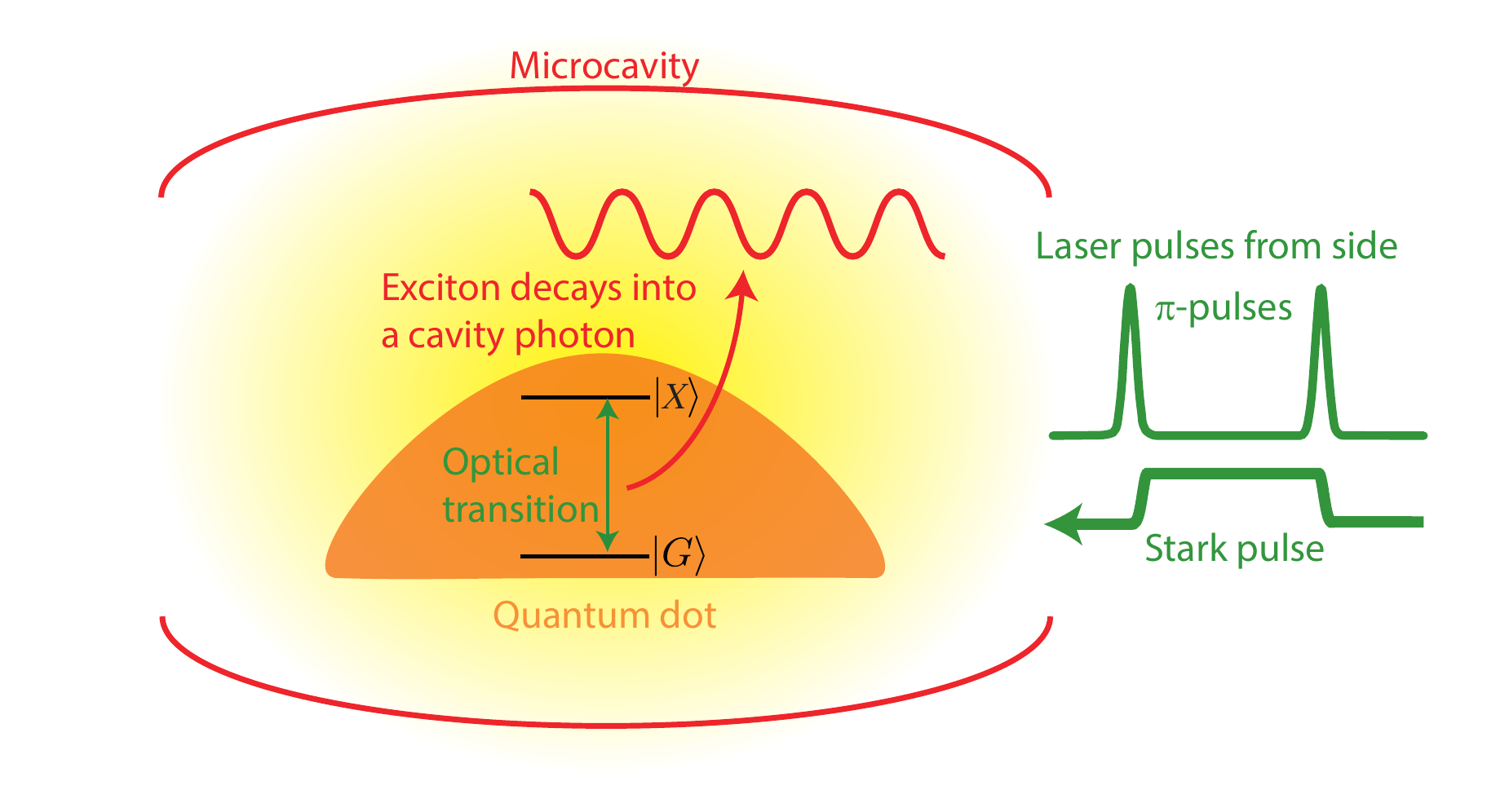}
	\caption{Sketch of the QDC system, where photons are created by recombination of the QD exciton. A Stark pulse from the side controls the photon emission such that Fock states can be generated on demand. }
	\label{fig:sketch}
\end{figure}

In the QDC case, there are several differences to the atomic situation. Firstly, the interaction between the QD and the cavity cannot be turned off by removing the QD from the cavity. Secondly, the transition frequency of the QD is set once the nanostructure is grown. Nonetheless, preparation schemes similar to those used in atomic cavities can be realized also in QDCs using the mechanisms sketched in Fig. \ref{fig:sketch}:
An interchange between exciting the QD with a sequence of $\pi$-pulses and controlling the effective cavity-QD coupling by inducing AC-Stark shifts lies at the heart of the two schemes we present here.
The difference between them is the way the AC-Stark pulses are used: in the first scheme, they lead to a cavity-QD coupling, while in the second one, they decouple the two subsystems at the end of the protocol.
Note that inducing ultrafast Stark shifts is also possible by electrical, rather than optical manipulation of the system \cite{Mukherjee2020}.

We start by assuming that the QD can be modelled as a two-level system, while we will discuss more realistic QD models in Sec.~\ref{sec:realistic}. This approximation holds very well for strongly confined charged QDs where the ground state is coupled to a trion state and higher excited states are energetically well separated. But it can also be realized in neutral QDs where, however, it entails constraints on the FSS, the polarizations of the driving laser, and the cavity modes.

The Hamiltonian for the two-level QD coupled to a single cavity mode and driven by external laser pulses reads:
\begin{align}
\label{eq:2LS}
H_{\t{2LS}}=\,&\hbar\w_X \XX\nn
+&\hbar\w_C a^\dagger a+\hbar g\left(a\s_X^\dagger+a^\dagger\s_X\right)\nn
-&\frac{\hbar}{2}\left(f_X^*(t)\s_X+f_X(t)\s_X^\dagger\right)\nn
f_X(t)=\,&f^{\t{pulses}}(t)+f^{\t{AC-Stark}}(t)
\end{align}
where $\X$ is the exciton state at energy $\hbar\w_{X}$ and $\s_X:=\,\GX$ is the operator for the transition between $\X$ and the ground state $\G$.
The energy of the latter is set to zero.
$a$ denotes the photonic annihilation operator.
The QDC is described by the Jaynes-Cummings model and the exciting and Stark laser pulses are represented by the function $f_X(t)$, which is specified in Appendix~\ref{app:hamiltonian}, in particular its two parts $f^{\t{pulses}}(t)$ and $f^{\t{AC-Stark}}(t)$. The cavity frequency is denoted by $\omega_C$ and its coupling to the QD by $g$.
We further account for the pure-dephasing type interaction with longitudinal acoustic (LA) phonons \cite{Besombes2001,Borri2001,Krummheuer2002,Axt2005,Reiter2019}, the radiative decay of the QD excitons, and cavity losses.
In this work, whenever we consider phononic effects, the phonons are assumed to be initially in thermal equilibrium at a temperature of $T=4\,$K. We solve the corresponding Liouville equation in a numerically complete manner by employing a path-integral formalism (for details see Refs.~\onlinecite{Vagov2011,Barth2016,Cygorek2017} and Appendix~\ref{app:hamiltonian}). The parameters for the calculation are given in Appendix~\ref{app:parameters}.

\subsection{Protocol with interrupted coupling - PIC}
\label{subsubsec:AC-Stark}
In a first step, we would like to translate the protocol known from atomic cavities as closely as possible to our solid state platform.
Therefore, we assume that the QD transition and the cavity mode are off-resonant. In particular, we assume that $\D\w_{\t{CX}}:=\w_{\t{C}}-\w_{\t{X}}>0$, i.e., the QD line lies below the cavity. In order to enable the efficient generation of a single photon in the cavity we apply an AC-Stark pulse tuned below the exciton line to bring the QD in resonance with the cavity. Each Stark pulse is a rectangular pulse with softened edges [cf., Eq.~\eqref{eq:Stark}]. Whilst in resonance, the QD exciton can emit a photon. Before a re-absorption of this photon occurs, we switch off the AC-Stark pulse, thus effectively interrupting the coupling of QD and cavity. Now these steps can be repeated to reach any desired photon state.
In the following, we refer to this scheme as \textit{protocol with interrupted coupling} (PIC).

\begin{figure}[t]
	\centering
	\includegraphics[width=0.45\textwidth]{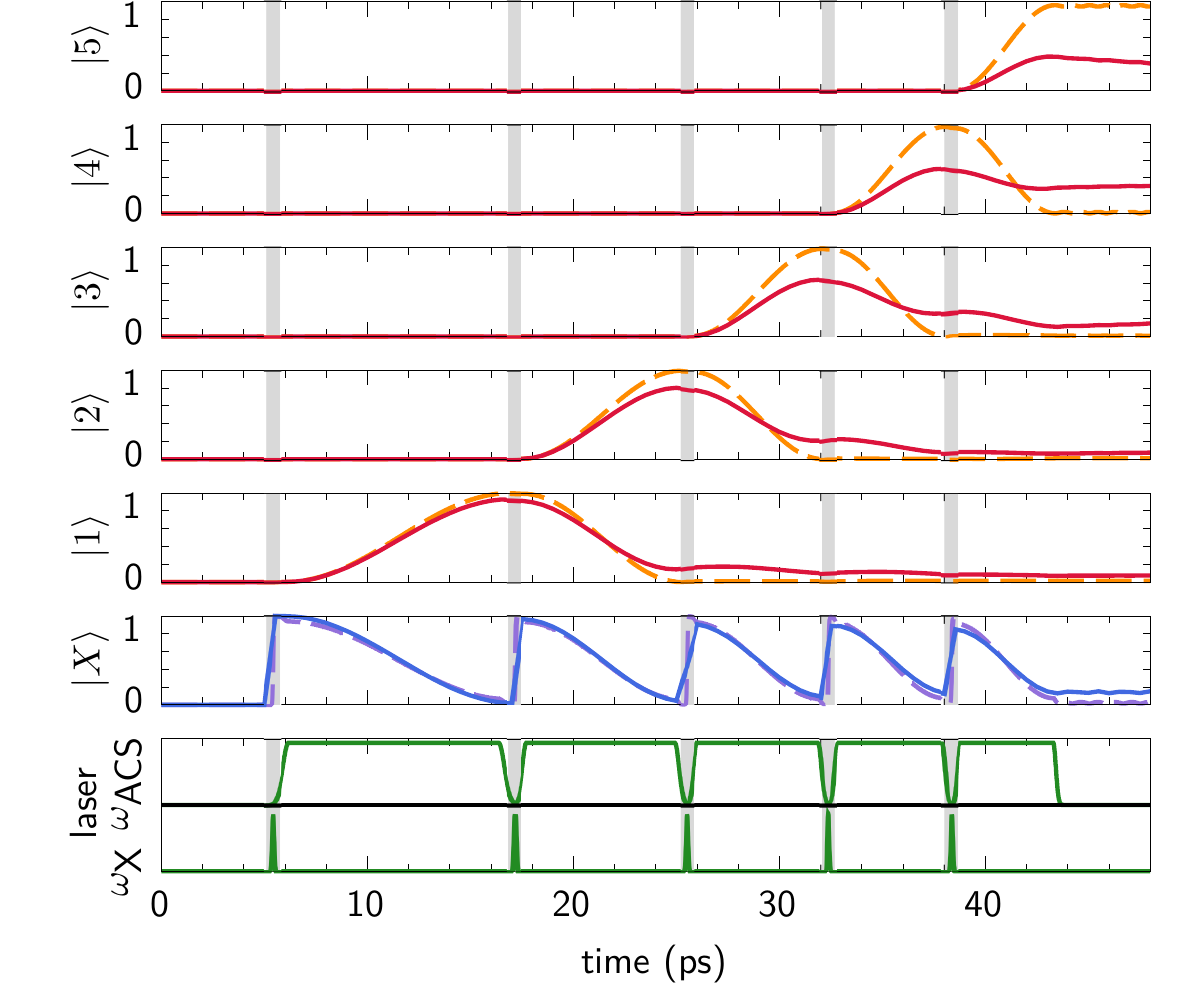}
	\caption{Dynamics of a QDC for the PIC. Panels from bottom to top: series of ultrafast $\pi$-pulses and off-resonant AC-Stark pulses (green);  the occupation of the exciton $|X\>$ (blue), occupation of the photon states $\vert n\> = \vert 1\>,\vert 2\>,...,\vert 5\>$ (red/orange). Dashed lines: without phonons and without losses. Solid lines with phonons as well as cavity losses and radiative decay.}
	\label{fig:AC_Stark}
\end{figure}

The corresponding dynamics of the PIC is displayed in Fig.~\ref{fig:AC_Stark}. The panels from bottom to top show the laser pulses (green), the occupation of the exciton $|X\>$ (blue), and the occupation of the photon states $\vert n\> = \vert 1\>,\vert 2\>,...,\vert 5\>$ (red/orange). We use our procedure of alternating $\pi$- and AC-Stark pulses five times until the Fock state $\vert 5\>$ is prepared. The dashed lines show the protocol in the ideal case of a two-level system without phonons and losses. Here, every Fock state is reached with a near-unity fidelity of $96.3\,\%$ (cf., Sec.~\ref{subsec:comparison} for a formal definition of this quantity). Because the Rabi frequency depends on the number of photons already present in the cavity, the length of the AC-Stark pulses for each step is reduced by $1/\sqrt{n}$ compared to the first Stark pulse. 

When taking both the phonon Hamiltonian and Markovian loss processes into account, both the exciton occupation and the occupation of the photon states are reduced (solid lines in Fig.~\ref{fig:AC_Stark}) and the fidelity of the protocol diminishes considerably. Nonetheless, we are still able to address each Fock state with our protocol. We note that cavity losses are responsible for the refilling of the Fock state with $n-1$ photons during the preparation of the state with $n$ leading to an additional loss of fidelity. Moreover, this effect yields second local maxima in the fidelity after the first ones intended by the preparation.

Even when considering all loss channels, we find a fidelity of 38.5~\%, which is in good comparison with other protocols \cite{Hofheinz2008,Tiedau2019}. A further advantage of the PIC is that the preparation is on-demand, which is a challenge in setups relying on parametric down-conversion. 

\subsection{Protocol with uninterrupted coupling - PUC}
\label{subsubsec:ultrashort}

\begin{figure}[t]
	\centering
	\includegraphics[width=0.45\textwidth]{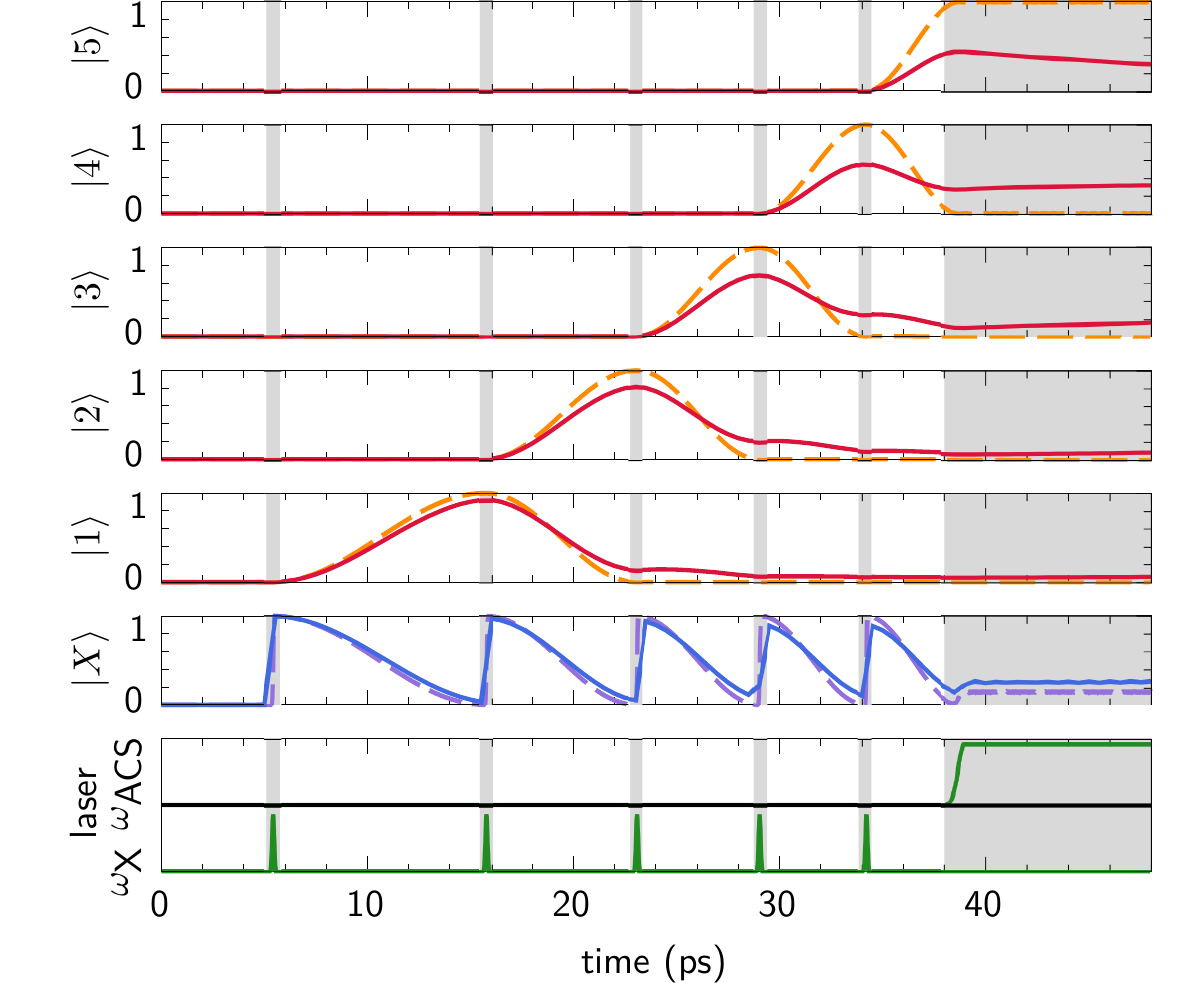}
	\caption{Dynamics of the QDC for the PUC. Panels and line types as in Fig.~\ref{fig:AC_Stark}}
	\label{fig:ultrashort}
\end{figure}

In contrast to the previous situation, now we consider the case that the QD and the cavity shall be grown such that they are on-resonance ($\D\w_{\t{CX}}=0$). Such systems are nowadays common in QDCs \cite{schneider2009single,reitzenstein2010quantum}. If now a series of resonant $\pi$-pulses hits the QD, such that the exciton performs half a Rabi oscillation in the time between the pulses, a number of photons according to the number of $\pi$-pulses is created.
Since in this scheme the coupling between QD and cavity is kept at resonance until the final state is reached, we refer to it as \textit{protocol with uninterrupted coupling} (PUC).
Note that like for the PIC the delay between the pulses has to be scaled by $1/\sqrt{n}$. Only when the target state has been reached, an AC-Stark pulse decouples the QD and the cavity to store the desired number state.

Figure~\ref{fig:ultrashort} shows the dynamics of the participating quantities for the PUC. In the phonon- and loss-free case (dashed lines in Fig.~\ref{fig:ultrashort}), the fidelity to the 5-photon state is $99.4\,\%$ which is even slightly higher than in the PIC, where we only reached $96.3\,\%$. Including the influence of phonons and losses, qualitatively the same effects as in the PIC can be observed, in particular the refilling of lower number states due to the cavity losses (solid lines in Fig.~\ref{fig:ultrashort}).
We achieve a fidelity including phonons and Markovian losses of $45,1\,\%$.

\subsection{Comparison of the two protocols}
\label{subsec:comparison}
\begin{figure}[t]
	\centering
	\includegraphics[width=\columnwidth]{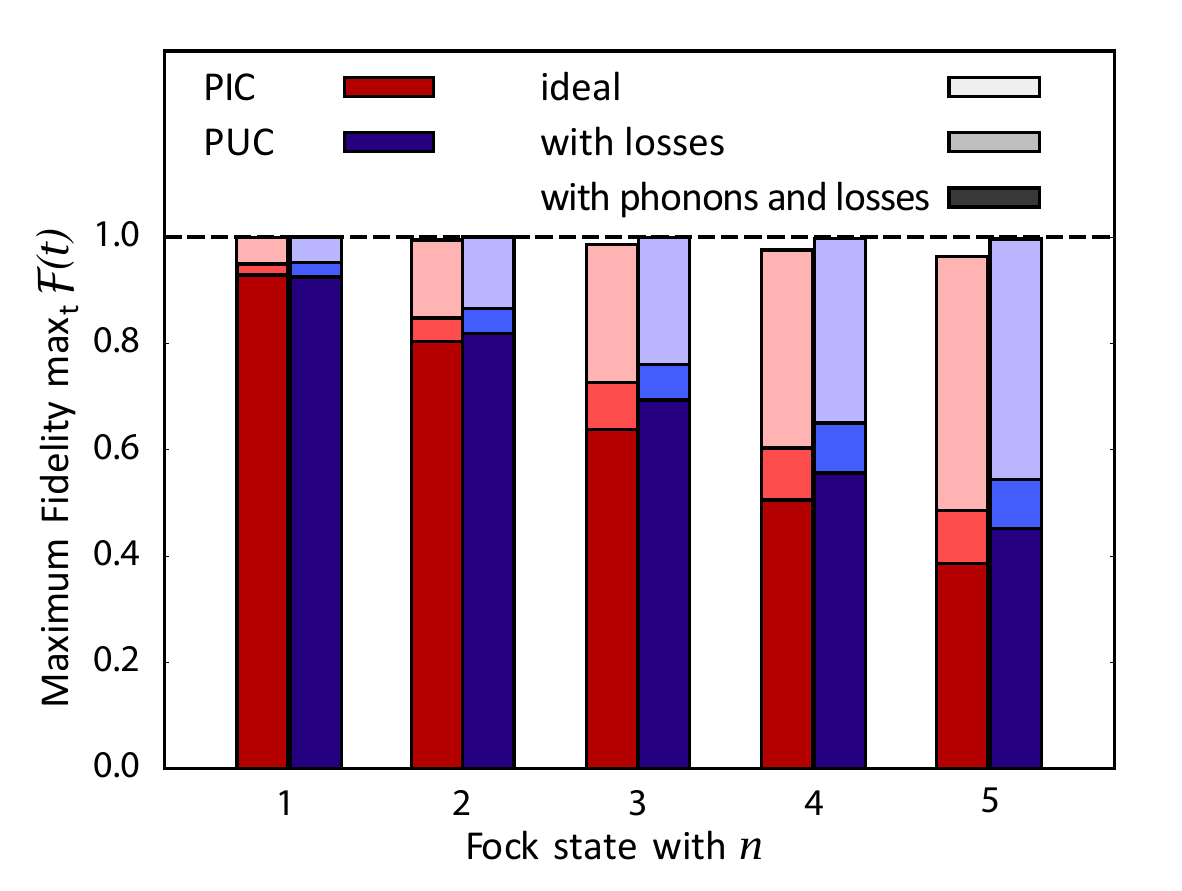}
	\caption{The maximum fidelity over time to the Fock states $\vert n\>$, $n\in\{1,2,...,5\}$ for the PIC (left bars, red) and PUC (right bars, blue) for the cases without phonons and without losses (light colors), without phonons and with losses (middle colors), and with phonons and with losses (dark colors).}
	\label{fig:hist_AC_vs_ultrashort}
\end{figure}
The striking difference between the two protocols is their total duration. The PUC is roughly $15\,\%$ faster than its PIC counterpart. This minimizes the time when losses can influence the dynamics.

We summarize our findings by looking at the maximum fidelity over time for each Fock state in Fig.~\ref{fig:hist_AC_vs_ultrashort} for the different levels of approximation. The fidelity is a generalization of the overlap projection between two pure states to mixed states first introduced by Jozsa \cite{Jozsa1994}. For two arbitrary density matrices $\r_1$ and $\r_2$ it is defined as
\begin{align}
\mathcal{F}(\r_1,\r_2)=\left[\Tr\left(\sqrt{\sqrt{\r_1}\r_2\sqrt{\r_1}}\right)\right]^2\, .
\end{align}
It is bounded between zero and unity and symmetric in its two arguments. In our case, we consider the photonic reduced density matrix $\r_1=\t{Tr}_{\t{QD}}\left[\t{Tr}_{\t{Ph}}\left(\r\right)\right]$, which is the full density matrix $\r$ obtained as a solution of Eq.~\eqref{eq:Liouville-von Neumann} traced over the phononic (Ph) and the electronic (QD) subspaces. We compare this state to the target Fock states $\vert n\>$, i.e., $\r_2=\vert n\>\<n\vert$. In this special case, the fidelity simplifies to the occupation of $\vert n\>$.

The red bars in Fig.~\ref{fig:hist_AC_vs_ultrashort} correspond to the PIC, while the blue bars are for the PUC. The lightly colored bars show the ideal case without phonons and without losses, the medium colored bars are without phonons, but with losses, and the dark colored bars are with phonons and with losses.

Looking at the fidelity of the Fock states, we can see several trends: The effect of cavity losses and radiative decay is more detrimental than the phonon influence. As an example, for the $n=5$-Fock state prepared by the PIC, the losses reduce the fidelity from $96.3\,\%$ to $48.5\,\%$, while the phonons further lower this value only to $38.5\,\%$. 
Note that this is a quantitative result for the considered GaAs/In(Ga)As QD at $T=4\,$K.
In particular, this behavior might change in other materials or at higher temperatures.
The PUC is better than the PIC for all cases. Overall, the $15\,\%$ saving of time in the PUC yields a clear benefit. The maximum fidelity to the $n=5$-Fock state including phonon and loss effects is now $45.1\,\%$ (compared with $38.5\,\%$ in the previous paragraph), a significant improvement by $17\,\%$.
Furthermore, the PUC poses less demand on the experimental realization, since only one AC-Stark pulse is necessary to decouple the system at the end of the protocol.
Even the field strength of this final pulse need not be precise, as required in the PIC, as long as it is large enough to effectively detune the dot from the cavity.

Therefore, we conclude that our protocols both perform well in comparison with existing protocols to prepare higher-order Fock states \cite{Hofheinz2008,Tiedau2019}.
The PUC outperforms the PIC with respect to the total duration as well as the fidelity as long as the conditions for using a two-level model are fulfilled.

\section{Protocols for a multi-level quantum dot system}
\label{sec:realistic}

For charged QDs, the transition between the residing electron and the trion state (i.e., the charged exciton) can be well modelled by a two-level system \cite{kaldewey2017demonstrating}. However, for a neutral QD, the assumption of a two-level system imposes further constraints. In particular, the single exciton manifold in a neutral QD comprises two states, which can be selectively addressed by circularly polarized laser pulses. A finite exchange interaction couples these states resulting in a finite FSS and corresponding new eigenstates that couple to linearly polarized light \cite{Axt2005,Bayer1999,Hoegele2004,Patton2003,Gammon1996}. Also, there exists the biexciton, which can be addressed using linearly polarized pulses. Additionally, using ultrashort $\pi$-pulses, might lead to the excitation of higher energetic exciton states.
Note that these states are also present in charged QDs, thus affecting the two-level approximation even in this favorable system.
In the following, we study how these deviations from a two-level system affect the Fock state preparation fidelity.

\subsection{Systems}
\label{sec:Model}

\begin{figure}[t]
	\centering
	\includegraphics[width=\columnwidth]{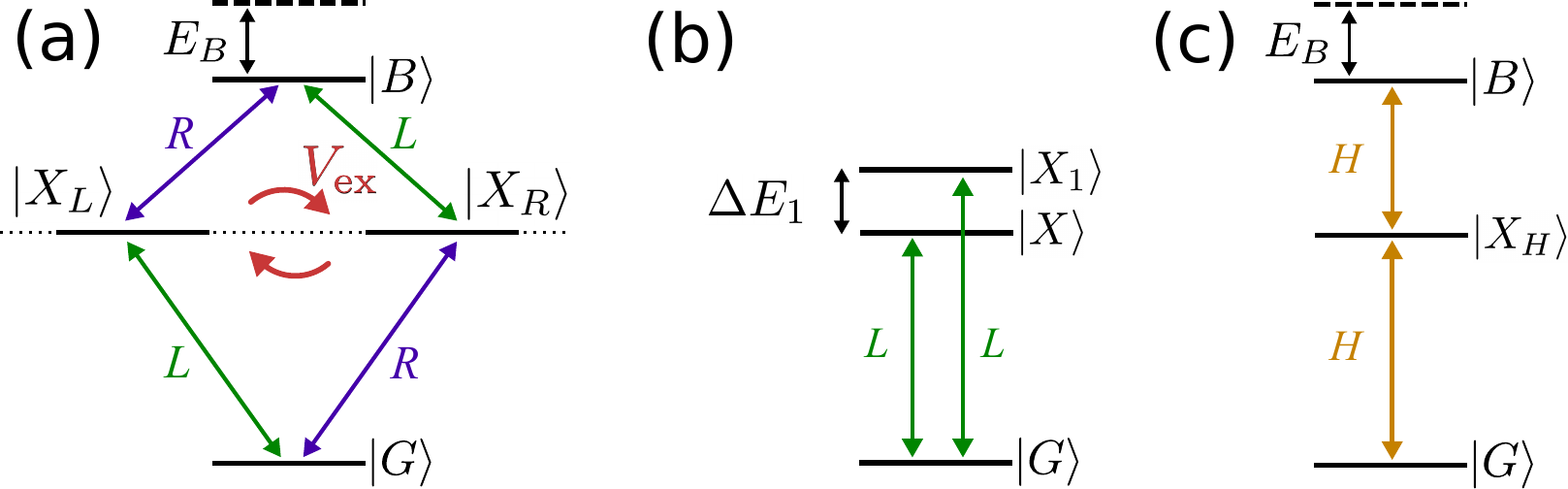}
	\caption{Level schemes for: (a) the four-level system representing a neutral QD, (b) the system with a higher energetic exciton and (c) the three-level system for a neutral QD driven by linearly polarized light. Arrows denote the optically allowed transitions with the corresponding polarization. }
	\label{fig:System}
\end{figure}

\begin{enumerate}
	\item \textbf{Four-level system (4LS)}
For modeling a neutral QD we consider a four-level system accounting for the ground state $\G$, the left and right circularly polarized exciton $\XL$ and $\XR$ as well as the biexciton $\B$
\begin{align}
\label{eq:4LS}
H_{\t{4LS}}=\,&\hbar\w_X\left(\XLXL+\XRXR\right)\nn
+&\hbar\frac{V_{\t{ex}}}{2}\left(\XLXR+\XRXL\right)\nn
+&\left(2\hbar\w_X - E_B\right)\BB\nn
+&\sum_{j=L,R}\left[\hbar\w_C a_j^\dagger a_j+\hbar g\left(a_j\s_j^\dagger+a_j^\dagger\s_j\right)\right]\nn
-&\frac{\hbar}{2}\left(f_L^*(t)\s_L+f_L(t)\s_L^\dagger\right)\,, 
\end{align}
$\hbar V_{\t{ex}}$ is the exchange splitting between the linearly polarized exciton states, and $E_B$ the biexciton binding energy. The allowed dipole selection rules lead to the following transition operator matrices
\begin{align}
\s_L:=\,&\GXL+\XRB\nn
\s_R:=\,&\GXR+\XLB\, .
\end{align}
The exchange interaction couples the two oppositely polarized exciton states $\XL$ and $\XR$, thus opening up a path to occupy the biexciton state $\B$ even when the QD is driven by pulses $f_L(t)$ that all have the same circular polarization. The cavity modes are described by the photon annihilation operators $a_L$ and $a_R$.
Note that here two cavity modes are coupled which are assumed to be degenerate.
The coupling strength is denoted by $g$ and assumed to be equal for all modes.
A sketch of the 4LS is shown in Fig.~\ref{fig:System}(a). 

\item \textbf{Higher energetic exciton system (HEES):}
To account for a higher energetic exciton, we assume a three-level system consisting of the ground state $\G$, the lowest energetic exciton $\X$, and an additional higher energetic exciton state $\XI$. 
\begin{align}
\label{eq:2LS+h}
H_{\t{HEES}}=&\,\hbar \w_X \XX \notag \\
	&+ \left(\hbar \w_X+\Delta E_{1}\right)\XIXI\nn
	+&\hbar\w_C a^\dagger a+\hbar g\left(a\s_X^\dagger+a^\dagger\s_X\right)\nn
-&\frac{\hbar}{2}\left(f_X^*(t)\s_X+f_X(t)\s_X^\dagger\right)\nn
-&\frac{\hbar}{2}\left(f_X^*(t)\s_1+f_X(t)\s_1^\dagger\right) \,, 
\end{align}
where the higher energetic exciton lies few tens of meV above the exciton energy as denoted by $\Delta E_1$. The corresponding transition operators in this system are
\begin{align}
\s_X:=\,&\GX\nn
\s_1:=\,&\GXI.
\end{align}
Because it is strongly off-resonant, we do not consider the coupling of $\s_1$ into the cavity mode.
A sketch of the HEES is shown in Fig.~\ref{fig:System}(b). 

\item \textbf{Three level system (3LS):}
The degeneracy of the two cavity modes in the 4LS implies that when driving with pulses all having the same linear polarization, one exciton becomes dark and the 4LS with two cavity modes reduces to a 3LS coupled only to a single linearly polarized mode as follows
\begin{align}
\label{eq:3LS}
H_{\t{3LS}}=&\,\hbar\w_{\tilde{X}} \XHXH \notag \\
	+&\left(2\hbar\w_{\tilde{X}} - E_B\right)\BB\nn
	+&\hbar\w_C a_H^\dagger a_H+\hbar g\left(a_H\s_H^\dagger+a_H^\dagger\s_H\right)\nn
-&\frac{\hbar}{2}\left(f_H^*(t)\s_H+f_H(t)\s_H^\dagger\right)\,, 
\end{align}
with the exciton energy lying at $\hbar\omega_{\tilde{X}}$ and the transition operators
\begin{align}
\s_H:=\, \GXH + \XHB\, .
\end{align}
The cavity photon is annihilated by the operator $a_H$ and the laser driving is described by the function $f_H(t)$.
A sketch of the 3LS is shown in Fig.~\ref{fig:System}(c). 

\end{enumerate}
For all systems, we again take into account the electron-phonon coupling and losses as described in Appendix~\ref{app:hamiltonian}.

\subsection{Results}
\begin{figure*}[t]
	\centering
	\includegraphics[width=0.75\textwidth]{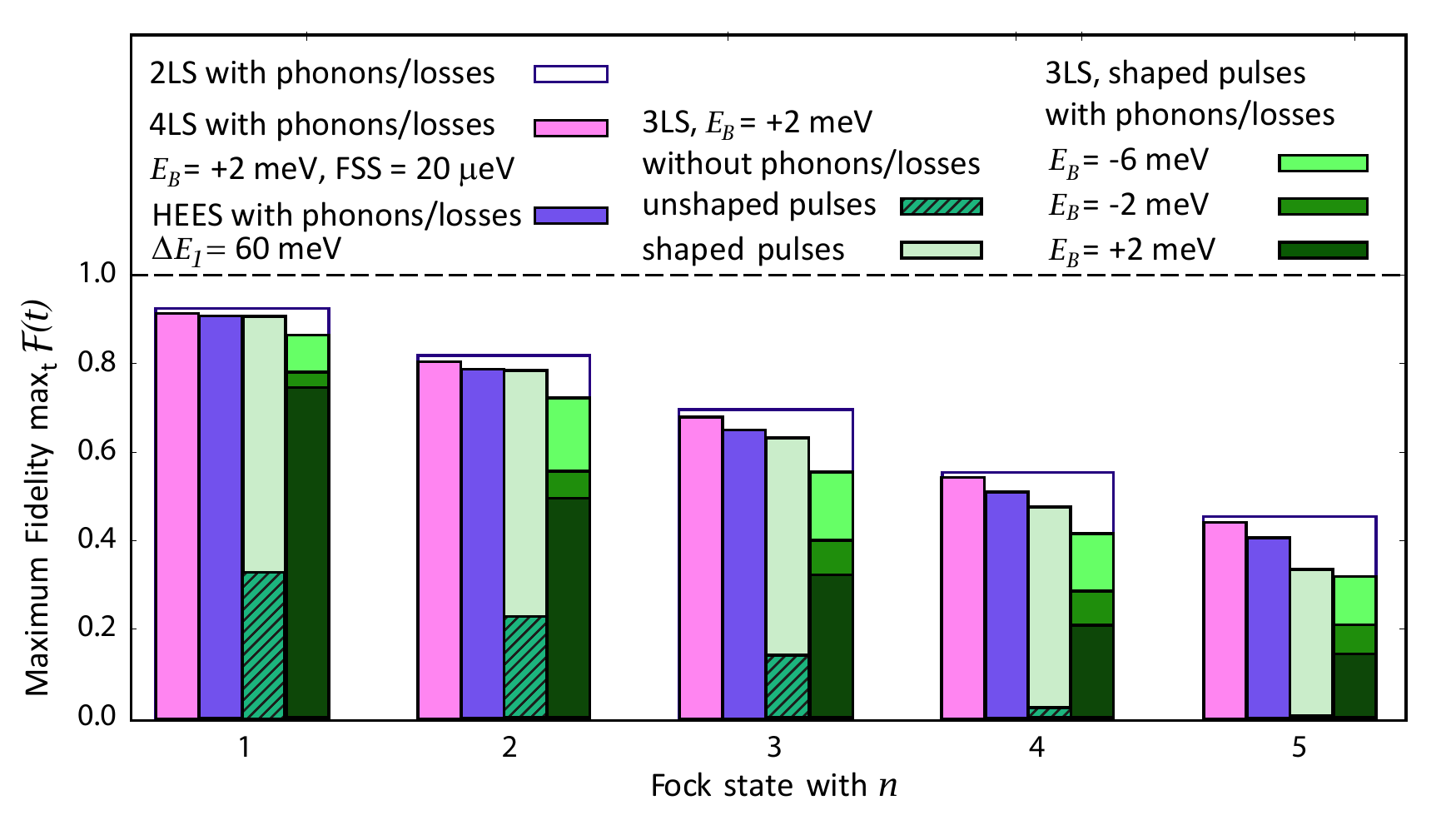}
	\caption{The maximum fidelity over time to the Fock states $\vert n\>$, $n\in\{1,2,...,5\}$ for different systems. The blue rectangles mark the values obtained for an ideal two-level system. From left to right: (pink) Four-level system with circular polarization with finite exchange splitting of $\hbar V_{\t{ex}}=20\,\mu$eV and a biexciton binding energy of $E_B=2\,$meV. (purple) Three-level system including an additional higher lying exciton state $\XI$ excited at an energetic difference of $\D E_1=60\,$meV. (green) Linearly excited biexcitonic three-level system using shaped or unshaped pulses (left) and shaped pulses for different biexciton binding energies $E_B=\pm 2\,$meV and $-6\,$meV (right). In all calculations the PUC has been used.
}
	\label{fig:fidelity}
\end{figure*}
Now we analyze the different influences on our preparation protocols. Figure~\ref{fig:fidelity} shows the maximal fidelity over time for the Fock states in the different systems. The values obtained for different $n$ are compared with the corresponding results of the PUC for an ideal two-level system accounting both for phonons and losses. This benchmark is displayed by a blue box around the bars.

\subsubsection{Influence of a finite exchange splitting - 4LS}
\label{subsubsec:V_ex}

The leftmost (pink) bars in Fig.~\ref{fig:fidelity} show the maximal fidelities for the four-level system with an exchange splitting of $\hbar V_{\t{ex}}=20\,\mu$eV and a biexciton binding energy of $E_B=2\,$meV, which represents values typically encountered in QDs. We find that for all Fock states the deterioration of the fidelity due to the finite exchange coupling is nonessential. As example consider the fidelity to the 5-photon Fock state, which reduces only to a value of $44.4\,\%$, i.e., only by $0.7\,\%$. For lower values of $V_{\t{ex}}$ the fidelity becomes even higher. We conclude that, typical exchange splittings of up to $20\,\mu$eV encountered in QDs do not influence the performance of the protocol significantly when it is excited with a well defined circular polarization.

\subsubsection{Influence of a higher energetic exciton state - HEES}
\label{subsubsec:X_1}

The purple bars in Fig.~\ref{fig:fidelity} show the resulting fidelities for exciting the three-level system with $\Delta E_{1}=60\,$meV, which is a typical value for strongly confined QDs \cite{Vagov2004}. Again, we find that the decrease of the fidelity is negligible. For $\Delta E_{1}=60\,$meV we obtain a 5-photon fidelity of $43.8\,\%$, while for $\Delta E_{1}=40\,$meV (not shown) the fidelity drops to $40.8\,\%$. This value is only $1.3\,\%$ (or $4.3\,\%$) below the result of the two-level model. Therefore, we conclude that the influence of the higher lying exciton state is not important as well as long as $\D E_1$ is sufficiently large which is the case for strongly confined QDs.

It is interesting to note that an increase of the pulse length does not necessarily improve the performance of the protocol in this case, even though this would result in a sharper spectral width of the pulse. To understand this effect, imagine lengthening the excitation pulses in the two-level case. When the pulses are long enough so that the dynamics induced by the cavity coupling $g$ sets in during the pulse, the exciton cannot reach its highest possible occupation anymore. Thus, the photon occupation and therefore the fidelity both decrease. Therefore, there is a competition between the detrimental influence of the simultaneous QD and cavity dynamics for longer pulses and the larger spectral width for shorter pulses which might lead to a spectral overlap with the higher lying exciton state for a given value of $\Delta E_{1}$.

\subsubsection{Linearly polarized excitation and pulse shaping - 3LS}
\label{subsec:3LS/B and Sham}

Finally, we study the 3LS representing a neutral QD driven by linearly polarized pulses. Here, $\B$ is the parasitic state with the difference that the energy spacing to be bridged is the biexciton binding energy $E_B$, which is an order of magnitude smaller than $\D E_1$.

This case is often studied in the literature as a limiting factor in the operation of QDs, both with respect to photonic properties \cite{Gustin2018} and to the preparation of specific QD states \cite{Gawarecki2012}.
Therefore, we investigate this rather unusual case of linearly polarized excitation in more detail.
Furthermore, this model is a prototype for a system with a parasitic state that is energetically close.
Discussing a solution in such a situation is a key point of this paragraph.

The green bars in Fig.~\ref{fig:fidelity} display the results for a finite binding energy. For $E_B=2\,$meV already in the phonon- and loss-free case (lower left green bars) the PUC acting on this biexcitonic 3LS breaks down drastically and the fidelity to the state $\vert 5\>$ drops below one percent. It is clear that the dynamics induced by significantly occupying the state $\B$ has a catastrophic effect on the success of the PUC.

To remedy this insufficiency, we employ shaped pulses that provide spectral holes at precisely the energies, where the parasitic states are found. A simple pulse shaping scheme for such purposes, proposed in Ref.~\onlinecite{Chen2001}, is based on a superposition of two Gaussian pulses with central frequencies separated by $E_B$ with different widths $s_1$ and $s_2$ in the time domain. The corresponding envelope function, which is put into Eq.~\eqref{eq:protocol}, is 
\begin{align}
\label{eq:shaped}
f_H^{\t{p}}(t)=\,&f_{H,0}\left(e^{-\left(\frac{t}{\sqrt{2}s_1}\right)^2}-e^{-\left(\frac{t}{\sqrt{2}s_2}\right)^2+i \frac{E_B}{\hbar} t}\right)
\end{align}
with $f_{H,0}=\left[\sqrt{\pi}\left[\sqrt{2}s_1-\sqrt{2}s_2 \exp{\left\lbrace-\left(\frac{E_B}{2\hbar}\sqrt{2}s_2\right)^2\right\rbrace}\right]\right]^{-1}$.
The two free parameters $s_1$ and $s_2$ can now be used to tune the spectral maximum to the transition to be addressed and the spectral hole to the parasitic state, in our case $\B$. For a binding energy of $E_B=2\,$meV this is achieved by setting $s_1=0.42\,$ps and $s_2=0.18\,$ps.
The spectrum of this shaped pulse as well as the respective spectra of its two constituent Gaussians are depicted in Fig.~\ref{fig:shaped} in the Appendix.

The upper left green bars in Fig.~\ref{fig:fidelity} show the results for the pulse shaping protocol without phonons and losses. Remarkably, this simple pulse shaping technique boosts the 5-photon fidelity from essentially zero back to $33.4\,\%$. 

Next we study how phonons and losses affect the PUC with shaped pulses (right green bars in Fig.~\ref{fig:fidelity}). We find a strong detrimental effect on the fidelity for $E_B=+2$~meV (lower right green bars), which reduces the fidelity to the 5-photon state from $33.4~\%$ to $14.4~\%$. 

We have also considered negative binding energies of $E_B=-2$~meV (middle right green bars) and $E_B=-6$~meV (upper right green bars).  QDs can be grown to have negative biexciton binding energies \cite{Dialynas2007}. Alternatively, $E_B$ can be tuned by applying electrostatic fields \cite{Ding2010,Trotta2012} or biaxial strain \cite{Trotta2013}, even to become negative.
Clearly, the fidelities to the Fock states become higher for negative $E_B$.
We notice that even larger negative binding energies give rise to higher fidelities.

A negative biexciton binding energy has the advantage that phonon emission processes that lead to an occupation of the biexciton state are suppressed since in a frame rotating with the laser frequency the biexciton state is not the energetically lowest one anymore \cite{Barth2016b}. Accordingly, we find for $E_B=-2$~meV that the fidelity is higher ($20.9~\%$) for $n=5$ than what is obtained for the corresponding positive binding energy $E_B=2\,$meV.
For the higher negative value of $E_B=-6\,$meV (upper right green bars in Fig.~\ref{fig:fidelity}) with the pulse shaping proposal (now with $s_1=0.21\,$ps and $s_2=0.04\,$ps) we obtain a 5-photon fidelity of $31.8\,\%$.

Even though phonons always degrade the performance of the here proposed protocols, they suppress an unfavorable process when the sign of the binding energy is chosen accordingly.
It is worthwhile to note that there are situations where phonons are even more beneficial \cite{Reiter2019}. Examples include phonon-assisted preparation schemes for excitons and biexcitons  \cite{Glaessl2013Pro,Reiter2014,Ardelt2014,Bounouar2015, Quilter2015,Reindl2017}, the introduction of off-resonant QD-cavity couplings  \cite{Ates2009,Naesby2008,Hohenester2009,Hohenester2010, Majumdar2011,Hughes2011,Florian2013,Calic2011,Laucht2011}, the phonon-induced enhancement of photon purities  \cite{Cosacchi2019} or the photon-pair entanglement \cite{Seidelmann2019} as well as enabling correlated emission from spatially remote QDs \cite{Reindl2017}.

In summary, even in the worst case of linearly polarized excitation pulses, the fidelity to the 5-photon Fock state can be enhanced from essentially zero to above $30\,\%$ even when the phonon-induced and other loss mechanisms are present. This is made possible by a combination of shaping the spectral characteristics of the laser pulses and tuning the biexciton binding energy to negative values, taking advantage of the otherwise interfering LA phonon coupling.

\section{Conclusion}
\label{sec:Conclusion}

We have presented and investigated two protocols for the preparation of higher-order Fock states in QDCs. To this end, we adapted a standard protocol, developed for the atomic physics platform, to QDC-based devices. The basic ingredients of this scheme are a series of $\pi$-pulse excitations and effective energy shifts induced by AC-Stark pulses that effectively interrupt the coupling between the QD and the cavity.
It turns out, however, that a protocol where the coupling is uninterrupted until the final target state is reached outperforms this standard scheme both in terms of duration and in terms of fidelity as long as it is justified to treat the system as a two-level system.
In our analysis, we include radiative decay, cavity losses, and phonon effects, which are specific to solid state QDCs. We predicted in the two-level system a fidelity to the Fock state $\vert 5\>$ of over $40\,\%$ when the protocol with uninterrupted coupling is used. This value is comparable to results achieved in superconducting qubit setups as well as by parametric down-conversion. We have tested our protocol against the influence of the fine structure splitting and higher excited exciton states and have demonstrated that in all of these cases fidelities above $40\,\%$ can be achieved. The advantage of using this protocol for a QDC platform is its total duration on a time scale of a few tens of picoseconds and its on-demand character.

We further discussed the excitation with linearly polarized pulses, which entails detrimental excitations of the biexciton in the QD. The coupling to the biexciton leads to a complete breakdown of our protocols already in the loss- and phonon-free case.  Nonetheless, a combination of a pulse shaping technique, tuning the biexciton binding energy to negative values, and the influence of phonons is able to push the fidelity to $\vert 5\>$ back to $31.8\,\%$.

With these easy to implement protocols, we are confident that also in solid state cavity systems the on-demand preparation of higher-order Fock states  on the picosecond time scale becomes possible.

\acknowledgments
M.Cy. thanks the Alexander-von-Humboldt foundation for support
through a Feodor Lynen fellowship. This work
was also funded by the Deutsche Forschungsgemeinschaft (DFG, German Research
Foundation) - project Nr. 419036043.

\appendix
\section{Theoretical Model}
\label{app:theory}

\subsection{Coupling Hamiltonian}
\label{app:hamiltonian}
The external laser pulses are described by
\begin{align}
\label{eq:protocol}
f_{j}(t)=\,&\underbrace{\sum_m f_j^{\t{p}}(t-t_m) e^{-i\w_{\t{p}}(t-t_m)}}_{:=f^{\t{pulses}}(t)}\nn
+&\underbrace{\sum_n f_j^{\t{ACS}}(t-t_n) e^{-i\w_{\t{ACS}}(t-t_n)}}_{:=f^{\t{AC-Stark}}(t)}\, .
\end{align}
$f_j^{\t{p}}(t)$ and $f_j^{\t{ACS}}(t)$ are the envelope functions of pump fields and AC-Stark pulses, respectively, with $j\in\{X,L,H\}$. $\w_{\t{p}}$ and $\w_{\t{ACS}}$ are the corresponding laser frequencies.
The pump fields are Gaussian $\pi$-pulses
\begin{align}
f_j^{\t{p}}(t)=\frac{\pi}{\sqrt{2\pi}\s}e^{-\frac{t^2}{2\s^2}}\,.
\end{align}
where $\s$ denotes the standard deviation, which is connected to the full width at half maximum (fwhm) by fwhm$=2\sqrt{2\ln{2}}\s$.
We assume $\w_{\t{p}}=\w_{X}$ for the 2LS, the HEES, and the 4LS and set $\w_{p}=\w_{\tilde{X}}$ for the 3LS, i.e., the laser is in resonance with an exciton resonance in the 2LS; the HEES, and the 3LS cases while it is in the middle between the fine-structure split exciton resonances for the 4LS.
The AC-Stark pulses are of rectangular shape with the edges smoothened by half Gaussians
\begin{align} \label{eq:Stark}
f_j^{\t{ACS}}(t)=
\begin{cases}
f_{j,s} e^{-\left(t+ \frac{\tau_{\t{length}}}{2}\right)^2/\left(2\s_{\t{on}}^2\right)} & t<-\frac{\tau_{\t{length}}}{2} \\
f_{j,s} & -\frac{\tau_{\t{length}}}{2}\leq t \leq \frac{\tau_{\t{length}}}{2} \\
f_{j,s} e^{-\left(t- \frac{\tau_{\t{length}}}{2}\right)^2/\left(2\s_{\t{off}}^2\right)} & t>\frac{\tau_{\t{length}}}{2}\, ,
\end{cases}
\end{align}
where $f_{j,s}$ denotes the field strength, i.e., the plateau height of the rectangular pulse, $\tau_{\t{length}}$ its length, and $\s_{\t{on}}$ ($\s_{\t{off}}$) the width of the rise (fall) of the smoothened edges.
Note that by letting $\s_{\t{on}}$ ($\s_{\t{off}}$) $\to 0$, high-frequency components disrupt the dynamics and thus the fidelity of the effective coupling of cavity and QD.
We checked this by observing a decreasing fidelity to the Fock state $\vert 1\>$ with lower values of $\s_{\t{on}}$ ($\s_{\t{off}}$).
We would like to stress that any smooth rise and fall of the rectangular pulse is sufficient, which we tested by using cosine edges instead.
The key point is the modeling of a realistic rectangular pulse, which has never mathematically precise Heaviside-shaped edges.

Also, the precise shape of the pump pulses is of minor importance as a test with hyperbolic secant pulses showed.
The main requirement is the shortness of the pulses compared with the time scale of the QDC dynamics.
Furthermore, the pulses need not be phase-locked, as we checked by introducing random mutual phases between the pulses.
This finding vastly reduces the experimental demand of realizing the proposed protocols.

\begin{figure}[t]
	\centering
	\includegraphics[width=\columnwidth]{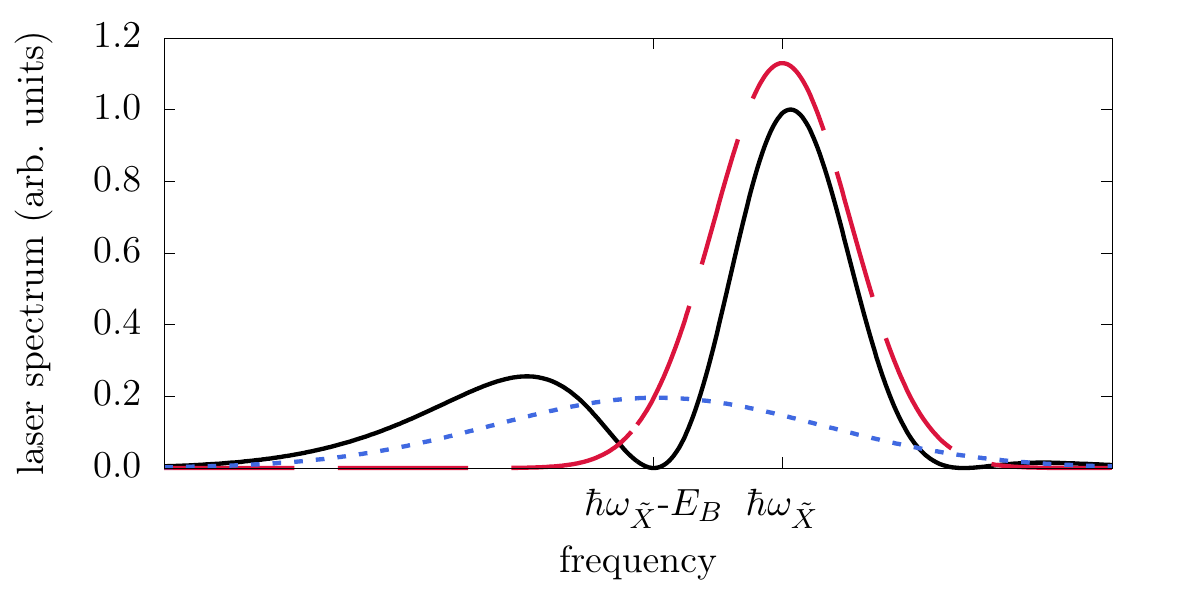}
	\caption{The spectrum of the shaped laser pulse as in Eq.~\eqref{eq:shaped} (black solid line) as well as the respective spectra of its two constituent Gaussians (red dashed and blue dotted lines). }
	\label{fig:shaped}
\end{figure}

In Paragraph~\ref{subsec:3LS/B and Sham}, the pump pulses $f_H^{\t{p}}(t)$ are shaped according to Eq.~\eqref{eq:shaped}.
The respective spectrum is depicted in Fig.~\ref{fig:shaped} as a black solid line.
The spectral hole at the biexciton binding energy $E_B$ is clearly visible.
The spectra of the constituent Gaussian pulses are plotted as red dashed and blue dotted lines.

The AC-Stark pulses are tuned below the exciton line by $\w_{\t{ACS,X}}:=\w_{\t{ACS}}-\w_{\t{X}}$ that is within the range of validity of the RWA. The resulting shift of the exciton line can be calculated from the energies of the laser dressed states.
Matching the shift to $\D\w_{\t{CX}}$, an AC-Stark pulse brings the exciton transition in resonance with the cavity provided that
\begin{align}
\label{eq:dCX'}
\D\w_{\t{CX}}=\D\w_{\t{ACS,X}}+\sqrt{\D\w_{\t{ACS,X}}^2+f_s^2}\, .
\end{align}
Note that tuning the coupling with an AC-Stark pulse is much more accurate than controlling the time of flight of an atom through a cavity. Any resonator has stray fields at its edges that depend on its geometry. Therefore, the time dependent coupling of the atom to the resonator is not rectangular, but has smoothened edges that are fixed by the geometry. In contrast, a laser pulse can be shaped to vary the edge characteristics, which introduces additional dials for optimizing the protocol.

The QD is coupled to LA phonons in a pure dephasing-type manner \cite{Besombes2001,Borri2001,Krummheuer2002,Axt2005,Reiter2019}. 
\begin{align}
H_{\t{Ph}}=\sum_\q\hbar\w_\q b_\q^\dagger b_\q
+\sum_{\q,\chi}n_\chi\left(\g_\q b_\q^\dagger+\g_\q^* b_\q\right)\vert\chi\>\<\chi\vert\, ,
\end{align}
$b_\q^\dagger$ and $b_\q$ are the phonon operators with wave vector $\q$ and energy $\hbar\w_\q$.
Bulk phonons with linear dispersion are considered that are coupled to the electronic states that are present in our respective systems $\vert\chi\>\in\{\XL,\XR,\XI,\B\}$ by the deformation potential-type coupling constant $\g_\q$. $n_\chi$ is the number of electron-hole pairs present in the state $\vert\chi\>$.

Finally, we take radiative recombination of the excitons and cavity loss processes into account by introducing Markovian Lindblad-type operators
\begin{align}
\mathcal{L}_{O,\Gamma}\bullet=\Gamma\left(O\bullet O\+ -\frac{1}{2}\left\lbrace\bullet,O\+ O\right\rbrace_+\right)\, ,
\end{align}
where $\{\cdot,\cdot\}_+$ denotes the anti-commutator.
$O$ is a system operator and $\Gamma$ the decay rate of the associated loss process.
We assume the radiative decay rate $\g$ is the same for all electronic transitions and take the same cavity loss rate $\kappa$ for both polarizations of the modes in the cavity.

The full Hamiltonian then reads as
\begin{align}
	H= H_{j} + H_{\text{Ph}}
\end{align}
with the different system Hamiltonians $H_j$ with $j\in\{\t{2LS, 4LS, HEES, 3LS}\}$ as defined in Sec.~\ref{sec:Model}. The dynamics of these systems is then described by the Liouville-von Neumann equation
\begin{align}
  \label{eq:Liouville-von Neumann}
  \frac{\partial}{\partial t} \r =\,&
-\frac{i}{\hbar}\{H,\r\}_-
+\mathcal{L}\r\, ,
\end{align}
where $\{\cdot,\cdot\}_-$ denotes the commutator.
The superoperator $\L\bullet$ comprises all Lindblad-type contributions to the dynamics for each considered system as follows:
\begin{widetext}
\[
\L\bullet=
\begin{cases}
\L_{a,\kappa}\bullet + \L_{\vert G\>\<X\vert,\g}\bullet & \t{for the 2LS}\\
\sum_{j=L,R}\left[\L_{a_j,\kappa}\bullet + \L_{\vert G\>\<X_j\vert,\g}\bullet + \L_{\vert X_j\>\<B\vert,\g}\bullet\right] & \t{for the 4LS}\\
\L_{a,\kappa}\bullet + \L_{\vert G\>\<X\vert,\g}\bullet + \L_{\vert G\>\<X_1\vert,\g}\bullet & \t{for the HEES}\\
\L_{a_H,\kappa}\bullet + \L_{\vert G\>\<X_H\vert,\g}\bullet + \L_{\vert X_H\>\<B\vert,\g}\bullet & \t{for the 3LS}\, .
\end{cases}
\]
\end{widetext}

We solve Eq.~\eqref{eq:Liouville-von Neumann} in a numerically complete manner by employing a path-integral formalism \cite{Makri1995a,Makri1995b,Vagov2011,Barth2016} that allows for the analytical integration of the infinitely many phonon modes.
Tracing the phonon degrees of freedom out yields a phonon induced memory kernel for the subsystem of interest.
By the term "numerically complete" we denote a solution that does not change noticeably by making the time discretization finer and the memory taken into account longer.
Recent advances within this method allows one to obtain solutions for systems with many quantum levels \cite{Cygorek2017}, which is paramount for the problem posed in this paper, since the relevant basis states to be considered are product states of the QD states and the number states of the two cavity modes.

\subsection{Parameters}
\label{app:parameters}

For the numerical calculations we use typical parameters for self-assembled strongly confined GaAs/In(Ga)As QDs \cite{Krummheuer2005,Cygorek2017}. The QD diameter is set to $6\,$nm. The cavity coupling is assumed to be $\hbar g=0.1\,$meV and the cavity losses are set to $\kappa=0.0085\,$ps$^{-1}$. Assuming a mode frequency of $\hbar\w_{\t{C}}=1.5\,$eV, this value of the loss rate corresponds to a cavity quality factor $Q\approx268,000$, which was reported in Ref.~\onlinecite{Schneider2016} as an extremely high but experimentally achievable value in QDCs. The radiative decay rate of the QD exciton is set to $\gamma=0.001\,$ps$^{-1}$. This corresponds to a typical lifetime of $1\,$ns. Whenever phonon effects are considered in this work, the phonons are assumed to be initially in thermal equilibrium at a temperature of $T=4\,$K.

The detuning between the cavity and the QD is assumed to be $\hbar\D\w_{\t{CX}}=5\,$meV in the case of the PIC and the difference between the AC-Stark pulse and the QD is set to $\hbar\D\w_{\t{ACS,X}}=-40\,$meV. Following from the condition Eq.~\eqref{eq:dCX'}, the AC-Stark amplitude has to be $\hbar f_s=21\,$meV. Furthermore, the width of the smoothened edges is chosen to be $\s_{\t{on}}=\s_{\t{off}}=0.28\,$ps. The pump pulses are on-resonance with the QD exciton and have a width of fwhm$=0.1\,$ps.

\subsection{Time discretization of the path-integral calculations}
\label{app:discretization}

The phonon induced memory kernel for GaAs/In(Ga)As QDs of $6\,$nm diameter at $T=4\,$K decays on a time scale of $\approx3\,$ps to zero \cite{Vagov2011,Barth2016,Cygorek2017}.
Therefore, numerically complete converged results are typically obtained for a discretization of $\D t_{\t{Ph}}=0.55\,$ps and $n_m=6$ memory steps.
With these values, the memory kernel is well sampled.
In this work, we use $\D t_{\t{Ph}}=0.5\,$ps and $n_m=7$.

To be able to resolve $0.1\,$ps-pulses, we first note that the dynamics induced by these ultrashort pulses is separated by one order of magnitude from the phonon time scale defined by the memory kernel.
Thus, on this fast time scale the phonon coupling has no influence on the system.
Therefore, a finer time discretization grid of $\D t=0.01\,$ps is put on top of the phonon discretization $\D t_{\t{Ph}}$.
On this finer grid $\D t$, the dynamics is calculated using the phonon-free propagator. This two-grid strategy is necessary since the discretization of the phonon memory with a time step of $\D t =0.01\,$ps would make the numerics intractable.

\begin{figure}[t]
	\centering
	\includegraphics[width=\columnwidth]{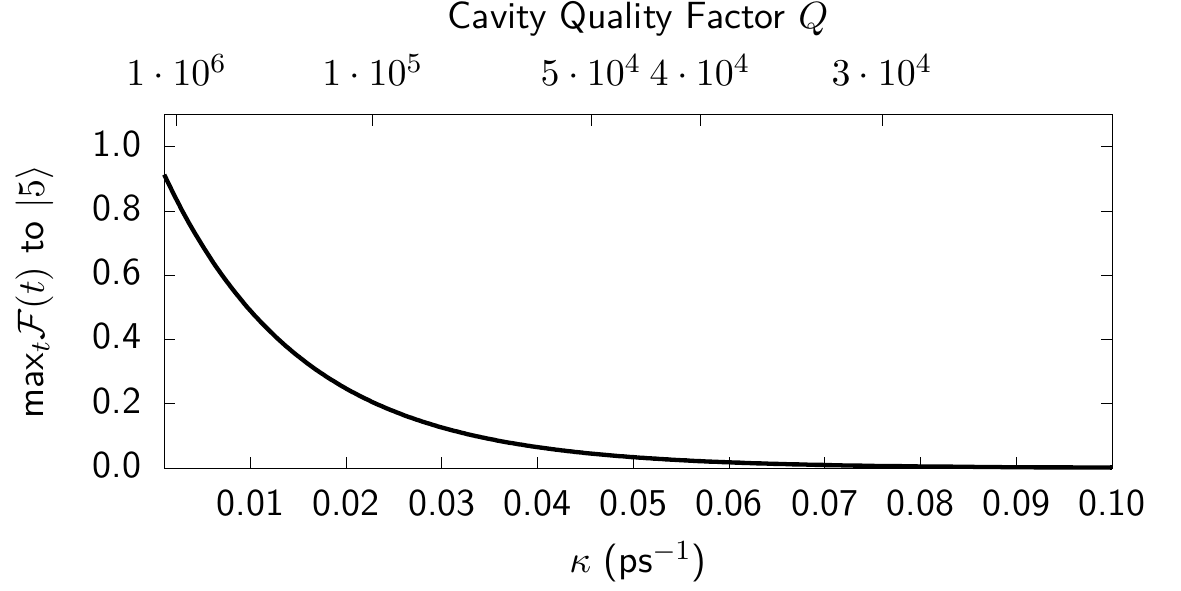}
	\caption{Dependence of the maximum fidelity over time to the 5-photon Fock state on the cavity loss rate. The cavity quality factor $Q$ is calculated assuming a mode frequency of $1.5\,$eV. The radiative decay has been kept fixed at $\gamma=0.001\,$ps$^{-1}$ (cf., Appendix~\ref{app:parameters}), while phonon effects are not considered in this plot.}
	\label{fig:kappa}
\end{figure}

\subsection{Implicit model for the PIC}
\label{app:implicit}

In the PIC presented in Paragraph~\ref{subsubsec:AC-Stark} for the two-level case, a sequence of AC-Stark pulses that are highly off-resonant ($\hbar\D\w_{\t{ACS,X}}=-40\,$meV) is the key ingredient.
To study the phonon effect on this protocol, the double grid scheme explained in Appendix~\ref{app:discretization} is not sufficient, since the fast oscillations induced by the AC-Stark pulses have to be sampled on the phonon discretization grid $\D t_{\t{Ph}}$ to fully capture the energy shifts induced by the off-resonant nature of the pulses and their interplay with the phonon environment.

To this end, an implicit model is introduced that does not include the AC-Stark pulse explicitly, i.e., $f^{\t{ACS}}(t)=0$ in the driving term $f(t)$.
Instead, in a frame co-rotating with the exciton frequency $\w_{X}$ the detuning $\D\w_{\t{CX}}$ becomes effectively time-dependent when a sequence of AC-Stark pulses is applied.
It is calculated according to Eq.~\eqref{eq:dCX'} by replacing $f_s$ with $f^{\t{ACS}}(t)$.
Thus, the largest detuning in the system is $\left|\hbar\D\w_{\t{CX}}\right|=5\,$meV in the implicit model compared with $\left|\hbar\D\w_{\t{ACS,X}}\right|=40\,$meV in the explicit incorporation of the AC-Stark pulses.

Comparing the implicit and the explicit model in the phonon-free case by studying the occupational dynamics of the system reveals that the amplitude of the oscillations is nearly identical, while their frequencies differ slightly for $\hbar\D\w_{\t{ACS,X}}\lesssim 10\,$meV.
For values greater than that and, in particular, in the limit $\D\w_{\t{ACS,X}}\to\infty$, the implicit and explicit models yield identical occupational dynamics.

\subsection{Influence of the cavity losses}
\label{app:cavity_losses}

The analysis in the main text, in particular, Sec.~\ref{subsec:comparison} and Fig.~\ref{fig:hist_AC_vs_ultrashort}, shows that the most influential parameter concerning the preparation fidelity of the Fock states is the cavity loss rate, i.e., the cavity quality factor $Q$.
Therefore, it is clear that lower Q values worsen and higher Q improve the results.
Nonetheless, it is of interest to quantify this effect due to its relevance for practical implementations of our scheme.
Fig.~\ref{fig:kappa} displays the fidelity reached for the 5-photon Fock state by the PUC as a function of the quality factor, which is varied by changing the cavity loss rate.

Indeed, the expectation that the fidelity to the 5-photon Fock state monotonically increases with higher Q is fulfilled.
The calculations shown are performed without considering phonon effects and keeping all the other parameters the same as in the main text.
In particular we have kept the radiative decay at the finite value of $\gamma=0.001\,$ps$^{-1}$.
This is the main reason why the fidelity stays noticeably below one in the limit of high Q values.

\bibliography{bib}

\begin{thebibliography}{78}%
\makeatletter
\providecommand \@ifxundefined [1]{%
 \@ifx{#1\undefined}
}%
\providecommand \@ifnum [1]{%
 \ifnum #1\expandafter \@firstoftwo
 \else \expandafter \@secondoftwo
 \fi
}%
\providecommand \@ifx [1]{%
 \ifx #1\expandafter \@firstoftwo
 \else \expandafter \@secondoftwo
 \fi
}%
\providecommand \natexlab [1]{#1}%
\providecommand \enquote  [1]{``#1''}%
\providecommand \bibnamefont  [1]{#1}%
\providecommand \bibfnamefont [1]{#1}%
\providecommand \citenamefont [1]{#1}%
\providecommand \href@noop [0]{\@secondoftwo}%
\providecommand \href [0]{\begingroup \@sanitize@url \@href}%
\providecommand \@href[1]{\@@startlink{#1}\@@href}%
\providecommand \@@href[1]{\endgroup#1\@@endlink}%
\providecommand \@sanitize@url [0]{\catcode `\\12\catcode `\$12\catcode
  `\&12\catcode `\#12\catcode `\^12\catcode `\_12\catcode `\%12\relax}%
\providecommand \@@startlink[1]{}%
\providecommand \@@endlink[0]{}%
\providecommand \url  [0]{\begingroup\@sanitize@url \@url }%
\providecommand \@url [1]{\endgroup\@href {#1}{\urlprefix }}%
\providecommand \urlprefix  [0]{URL }%
\providecommand \Eprint [0]{\href }%
\providecommand \doibase [0]{http://dx.doi.org/}%
\providecommand \selectlanguage [0]{\@gobble}%
\providecommand \bibinfo  [0]{\@secondoftwo}%
\providecommand \bibfield  [0]{\@secondoftwo}%
\providecommand \translation [1]{[#1]}%
\providecommand \BibitemOpen [0]{}%
\providecommand \bibitemStop [0]{}%
\providecommand \bibitemNoStop [0]{.\EOS\space}%
\providecommand \EOS [0]{\spacefactor3000\relax}%
\providecommand \BibitemShut  [1]{\csname bibitem#1\endcsname}%
\let\auto@bib@innerbib\@empty
\bibitem [{\citenamefont {Michler}\ \emph {et~al.}(2000)\citenamefont
  {Michler}, \citenamefont {Kiraz}, \citenamefont {Becher}, \citenamefont
  {Schoenfeld}, \citenamefont {Petroff}, \citenamefont {Zhang}, \citenamefont
  {Hu},\ and\ \citenamefont {Imamoglu}}]{Michler2000}%
  \BibitemOpen
  \bibfield  {author} {\bibinfo {author} {\bibfnamefont {P.}~\bibnamefont
  {Michler}}, \bibinfo {author} {\bibfnamefont {A.}~\bibnamefont {Kiraz}},
  \bibinfo {author} {\bibfnamefont {C.}~\bibnamefont {Becher}}, \bibinfo
  {author} {\bibfnamefont {W.~V.}\ \bibnamefont {Schoenfeld}}, \bibinfo
  {author} {\bibfnamefont {P.~M.}\ \bibnamefont {Petroff}}, \bibinfo {author}
  {\bibfnamefont {L.}~\bibnamefont {Zhang}}, \bibinfo {author} {\bibfnamefont
  {E.}~\bibnamefont {Hu}}, \ and\ \bibinfo {author} {\bibfnamefont
  {A.}~\bibnamefont {Imamoglu}},\ }\href {\doibase
  10.1126/science.290.5500.2282} {\bibfield  {journal} {\bibinfo  {journal}
  {Science}\ }\textbf {\bibinfo {volume} {290}},\ \bibinfo {pages} {2282}
  (\bibinfo {year} {2000})}\BibitemShut {NoStop}%
\bibitem [{\citenamefont {Santori}\ \emph {et~al.}(2001)\citenamefont
  {Santori}, \citenamefont {Pelton}, \citenamefont {Solomon}, \citenamefont
  {Dale},\ and\ \citenamefont {Yamamoto}}]{Santori2001}%
  \BibitemOpen
  \bibfield  {author} {\bibinfo {author} {\bibfnamefont {C.}~\bibnamefont
  {Santori}}, \bibinfo {author} {\bibfnamefont {M.}~\bibnamefont {Pelton}},
  \bibinfo {author} {\bibfnamefont {G.}~\bibnamefont {Solomon}}, \bibinfo
  {author} {\bibfnamefont {Y.}~\bibnamefont {Dale}}, \ and\ \bibinfo {author}
  {\bibfnamefont {Y.}~\bibnamefont {Yamamoto}},\ }\href
  {https://link.aps.org/doi/10.1103/PhysRevLett.86.1502} {\bibfield  {journal}
  {\bibinfo  {journal} {Phys.\ Rev.\ Lett.}\ }\textbf {\bibinfo {volume}
  {86}},\ \bibinfo {pages} {1502} (\bibinfo {year} {2001})}\BibitemShut
  {NoStop}%
\bibitem [{\citenamefont {Santori}\ \emph {et~al.}(2002)\citenamefont
  {Santori}, \citenamefont {Fattal}, \citenamefont {Vuckovic}, \citenamefont
  {Solomon},\ and\ \citenamefont {Yamamoto}}]{Santori2002}%
  \BibitemOpen
  \bibfield  {author} {\bibinfo {author} {\bibfnamefont {C.}~\bibnamefont
  {Santori}}, \bibinfo {author} {\bibfnamefont {D.}~\bibnamefont {Fattal}},
  \bibinfo {author} {\bibfnamefont {J.}~\bibnamefont {Vuckovic}}, \bibinfo
  {author} {\bibfnamefont {G.~S.}\ \bibnamefont {Solomon}}, \ and\ \bibinfo
  {author} {\bibfnamefont {Y.}~\bibnamefont {Yamamoto}},\ }\href
  {http://dx.doi.org/10.1038/nature01086} {\bibfield  {journal} {\bibinfo
  {journal} {Nature}\ }\textbf {\bibinfo {volume} {419}},\ \bibinfo {pages}
  {594} (\bibinfo {year} {2002})}\BibitemShut {NoStop}%
\bibitem [{\citenamefont {He}\ \emph {et~al.}(2013)\citenamefont {He},
  \citenamefont {He}, \citenamefont {Wei}, \citenamefont {Wu}, \citenamefont
  {Atat{\"u}re}, \citenamefont {Schneider}, \citenamefont {H{\"o}fling},
  \citenamefont {Kamp}, \citenamefont {Lu},\ and\ \citenamefont
  {Pan}}]{He2013}%
  \BibitemOpen
  \bibfield  {author} {\bibinfo {author} {\bibfnamefont {Y.-M.}\ \bibnamefont
  {He}}, \bibinfo {author} {\bibfnamefont {Y.}~\bibnamefont {He}}, \bibinfo
  {author} {\bibfnamefont {Y.-J.}\ \bibnamefont {Wei}}, \bibinfo {author}
  {\bibfnamefont {D.}~\bibnamefont {Wu}}, \bibinfo {author} {\bibfnamefont
  {M.}~\bibnamefont {Atat{\"u}re}}, \bibinfo {author} {\bibfnamefont
  {C.}~\bibnamefont {Schneider}}, \bibinfo {author} {\bibfnamefont
  {S.}~\bibnamefont {H{\"o}fling}}, \bibinfo {author} {\bibfnamefont
  {M.}~\bibnamefont {Kamp}}, \bibinfo {author} {\bibfnamefont {C.-Y.}\
  \bibnamefont {Lu}}, \ and\ \bibinfo {author} {\bibfnamefont {J.-W.}\
  \bibnamefont {Pan}},\ }\href {http://dx.doi.org/10.1038/nnano.2012.262}
  {\bibfield  {journal} {\bibinfo  {journal} {Nat. Nanotechnol.}\ }\textbf
  {\bibinfo {volume} {8}},\ \bibinfo {pages} {213} (\bibinfo {year}
  {2013})}\BibitemShut {NoStop}%
\bibitem [{\citenamefont {Wei}\ \emph {et~al.}(2014)\citenamefont {Wei},
  \citenamefont {He}, \citenamefont {Chen}, \citenamefont {Hu}, \citenamefont
  {He}, \citenamefont {Wu}, \citenamefont {Schneider}, \citenamefont {Kamp},
  \citenamefont {H{\"o}fling}, \citenamefont {Lu},\ and\ \citenamefont
  {Pan}}]{Wei2014Det}%
  \BibitemOpen
  \bibfield  {author} {\bibinfo {author} {\bibfnamefont {Y.-J.}\ \bibnamefont
  {Wei}}, \bibinfo {author} {\bibfnamefont {Y.-M.}\ \bibnamefont {He}},
  \bibinfo {author} {\bibfnamefont {M.-C.}\ \bibnamefont {Chen}}, \bibinfo
  {author} {\bibfnamefont {Y.-N.}\ \bibnamefont {Hu}}, \bibinfo {author}
  {\bibfnamefont {Y.}~\bibnamefont {He}}, \bibinfo {author} {\bibfnamefont
  {D.}~\bibnamefont {Wu}}, \bibinfo {author} {\bibfnamefont {C.}~\bibnamefont
  {Schneider}}, \bibinfo {author} {\bibfnamefont {M.}~\bibnamefont {Kamp}},
  \bibinfo {author} {\bibfnamefont {S.}~\bibnamefont {H{\"o}fling}}, \bibinfo
  {author} {\bibfnamefont {C.-Y.}\ \bibnamefont {Lu}}, \ and\ \bibinfo {author}
  {\bibfnamefont {J.-W.}\ \bibnamefont {Pan}},\ }\href
  {https://doi.org/10.1021/nl503081n} {\bibfield  {journal} {\bibinfo
  {journal} {Nano\ Lett.}\ }\textbf {\bibinfo {volume} {14}},\ \bibinfo {pages}
  {6515} (\bibinfo {year} {2014})}\BibitemShut {NoStop}%
\bibitem [{\citenamefont {Ding}\ \emph {et~al.}(2016)\citenamefont {Ding},
  \citenamefont {He}, \citenamefont {Duan}, \citenamefont {Gregersen},
  \citenamefont {Chen}, \citenamefont {Unsleber}, \citenamefont {Maier},
  \citenamefont {Schneider}, \citenamefont {Kamp}, \citenamefont {H{\"o}fling},
  \citenamefont {Lu},\ and\ \citenamefont {Pan}}]{Ding2016}%
  \BibitemOpen
  \bibfield  {author} {\bibinfo {author} {\bibfnamefont {X.}~\bibnamefont
  {Ding}}, \bibinfo {author} {\bibfnamefont {Y.}~\bibnamefont {He}}, \bibinfo
  {author} {\bibfnamefont {Z.-C.}\ \bibnamefont {Duan}}, \bibinfo {author}
  {\bibfnamefont {N.}~\bibnamefont {Gregersen}}, \bibinfo {author}
  {\bibfnamefont {M.-C.}\ \bibnamefont {Chen}}, \bibinfo {author}
  {\bibfnamefont {S.}~\bibnamefont {Unsleber}}, \bibinfo {author}
  {\bibfnamefont {S.}~\bibnamefont {Maier}}, \bibinfo {author} {\bibfnamefont
  {C.}~\bibnamefont {Schneider}}, \bibinfo {author} {\bibfnamefont
  {M.}~\bibnamefont {Kamp}}, \bibinfo {author} {\bibfnamefont {S.}~\bibnamefont
  {H{\"o}fling}}, \bibinfo {author} {\bibfnamefont {C.-Y.}\ \bibnamefont {Lu}},
  \ and\ \bibinfo {author} {\bibfnamefont {J.-W.}\ \bibnamefont {Pan}},\ }\href
  {\doibase 10.1103/PhysRevLett.116.020401} {\bibfield  {journal} {\bibinfo
  {journal} {Phys. Rev. Lett.}\ }\textbf {\bibinfo {volume} {116}},\ \bibinfo
  {pages} {020401} (\bibinfo {year} {2016})}\BibitemShut {NoStop}%
\bibitem [{\citenamefont {Somaschi}\ \emph {et~al.}(2016)\citenamefont
  {Somaschi}, \citenamefont {Giesz}, \citenamefont {De~Santis}, \citenamefont
  {Loredo}, \citenamefont {Almeida}, \citenamefont {Hornecker}, \citenamefont
  {Portalupi}, \citenamefont {Grange}, \citenamefont {Ant{\'o}n}, \citenamefont
  {Demory}, \citenamefont {G{\'o}mez}, \citenamefont {Sagnes}, \citenamefont
  {Lanzillotti-Kimura}, \citenamefont {Lema{\'i}tre}, \citenamefont {Auffeves},
  \citenamefont {White}, \citenamefont {Lanco},\ and\ \citenamefont
  {Senellart}}]{Somaschi2016}%
  \BibitemOpen
  \bibfield  {author} {\bibinfo {author} {\bibfnamefont {N.}~\bibnamefont
  {Somaschi}}, \bibinfo {author} {\bibfnamefont {V.}~\bibnamefont {Giesz}},
  \bibinfo {author} {\bibfnamefont {L.}~\bibnamefont {De~Santis}}, \bibinfo
  {author} {\bibfnamefont {J.~C.}\ \bibnamefont {Loredo}}, \bibinfo {author}
  {\bibfnamefont {M.~P.}\ \bibnamefont {Almeida}}, \bibinfo {author}
  {\bibfnamefont {G.}~\bibnamefont {Hornecker}}, \bibinfo {author}
  {\bibfnamefont {S.~L.}\ \bibnamefont {Portalupi}}, \bibinfo {author}
  {\bibfnamefont {T.}~\bibnamefont {Grange}}, \bibinfo {author} {\bibfnamefont
  {C.}~\bibnamefont {Ant{\'o}n}}, \bibinfo {author} {\bibfnamefont
  {J.}~\bibnamefont {Demory}}, \bibinfo {author} {\bibfnamefont
  {C.}~\bibnamefont {G{\'o}mez}}, \bibinfo {author} {\bibfnamefont
  {I.}~\bibnamefont {Sagnes}}, \bibinfo {author} {\bibfnamefont {N.~D.}\
  \bibnamefont {Lanzillotti-Kimura}}, \bibinfo {author} {\bibfnamefont
  {A.}~\bibnamefont {Lema{\'i}tre}}, \bibinfo {author} {\bibfnamefont
  {A.}~\bibnamefont {Auffeves}}, \bibinfo {author} {\bibfnamefont {A.~G.}\
  \bibnamefont {White}}, \bibinfo {author} {\bibfnamefont {L.}~\bibnamefont
  {Lanco}}, \ and\ \bibinfo {author} {\bibfnamefont {P.}~\bibnamefont
  {Senellart}},\ }\href {http://dx.doi.org/10.1038/nphoton.2016.23} {\bibfield
  {journal} {\bibinfo  {journal} {Nat. Photonics}\ }\textbf {\bibinfo {volume}
  {10}},\ \bibinfo {pages} {340} (\bibinfo {year} {2016})}\BibitemShut
  {NoStop}%
\bibitem [{\citenamefont {Schweickert}\ \emph {et~al.}(2018)\citenamefont
  {Schweickert}, \citenamefont {J{\"o}ns}, \citenamefont {Zeuner},
  \citenamefont {Covre~da Silva}, \citenamefont {Huang}, \citenamefont
  {Lettner}, \citenamefont {Reindl}, \citenamefont {Zichi}, \citenamefont
  {Trotta}, \citenamefont {Rastelli},\ and\ \citenamefont
  {Zwiller}}]{Schweickert2018}%
  \BibitemOpen
  \bibfield  {author} {\bibinfo {author} {\bibfnamefont {L.}~\bibnamefont
  {Schweickert}}, \bibinfo {author} {\bibfnamefont {K.~D.}\ \bibnamefont
  {J{\"o}ns}}, \bibinfo {author} {\bibfnamefont {K.~D.}\ \bibnamefont
  {Zeuner}}, \bibinfo {author} {\bibfnamefont {S.~F.}\ \bibnamefont {Covre~da
  Silva}}, \bibinfo {author} {\bibfnamefont {H.}~\bibnamefont {Huang}},
  \bibinfo {author} {\bibfnamefont {T.}~\bibnamefont {Lettner}}, \bibinfo
  {author} {\bibfnamefont {M.}~\bibnamefont {Reindl}}, \bibinfo {author}
  {\bibfnamefont {J.}~\bibnamefont {Zichi}}, \bibinfo {author} {\bibfnamefont
  {R.}~\bibnamefont {Trotta}}, \bibinfo {author} {\bibfnamefont
  {A.}~\bibnamefont {Rastelli}}, \ and\ \bibinfo {author} {\bibfnamefont
  {V.}~\bibnamefont {Zwiller}},\ }\href {\doibase 10.1063/1.5020038} {\bibfield
   {journal} {\bibinfo  {journal} {Appl. Phys. Lett.}\ }\textbf {\bibinfo
  {volume} {112}},\ \bibinfo {pages} {093106} (\bibinfo {year}
  {2018})}\BibitemShut {NoStop}%
\bibitem [{\citenamefont {Hanschke}\ \emph {et~al.}(2018)\citenamefont
  {Hanschke}, \citenamefont {Fischer}, \citenamefont {Appel}, \citenamefont
  {Lukin}, \citenamefont {Wierzbowski}, \citenamefont {Sun}, \citenamefont
  {Trivedi}, \citenamefont {Vu\v{c}kovi\'{c}}, \citenamefont {Finley},\ and\
  \citenamefont {M\"uller}}]{Hanschke2018}%
  \BibitemOpen
  \bibfield  {author} {\bibinfo {author} {\bibfnamefont {L.}~\bibnamefont
  {Hanschke}}, \bibinfo {author} {\bibfnamefont {K.~A.}\ \bibnamefont
  {Fischer}}, \bibinfo {author} {\bibfnamefont {S.}~\bibnamefont {Appel}},
  \bibinfo {author} {\bibfnamefont {D.}~\bibnamefont {Lukin}}, \bibinfo
  {author} {\bibfnamefont {J.}~\bibnamefont {Wierzbowski}}, \bibinfo {author}
  {\bibfnamefont {S.}~\bibnamefont {Sun}}, \bibinfo {author} {\bibfnamefont
  {R.}~\bibnamefont {Trivedi}}, \bibinfo {author} {\bibfnamefont
  {J.}~\bibnamefont {Vu\v{c}kovi\'{c}}}, \bibinfo {author} {\bibfnamefont
  {J.~J.}\ \bibnamefont {Finley}}, \ and\ \bibinfo {author} {\bibfnamefont
  {K.}~\bibnamefont {M\"uller}},\ }\href
  {http://www.nature.com/articles/s41534-018-0092-0} {\bibfield  {journal}
  {\bibinfo  {journal} {npj Quantum Inf.}\ }\textbf {\bibinfo {volume} {4}}
  (\bibinfo {year} {2018})}\BibitemShut {NoStop}%
\bibitem [{\citenamefont {Cosacchi}\ \emph {et~al.}(2019)\citenamefont
  {Cosacchi}, \citenamefont {Ungar}, \citenamefont {Cygorek}, \citenamefont
  {Vagov},\ and\ \citenamefont {Axt}}]{Cosacchi2019}%
  \BibitemOpen
  \bibfield  {author} {\bibinfo {author} {\bibfnamefont {M.}~\bibnamefont
  {Cosacchi}}, \bibinfo {author} {\bibfnamefont {F.}~\bibnamefont {Ungar}},
  \bibinfo {author} {\bibfnamefont {M.}~\bibnamefont {Cygorek}}, \bibinfo
  {author} {\bibfnamefont {A.}~\bibnamefont {Vagov}}, \ and\ \bibinfo {author}
  {\bibfnamefont {V.~M.}\ \bibnamefont {Axt}},\ }\href {\doibase
  10.1103/PhysRevLett.123.017403} {\bibfield  {journal} {\bibinfo  {journal}
  {Phys. Rev. Lett.}\ }\textbf {\bibinfo {volume} {123}},\ \bibinfo {pages}
  {017403} (\bibinfo {year} {2019})}\BibitemShut {NoStop}%
\bibitem [{\citenamefont {Akopian}\ \emph {et~al.}(2006)\citenamefont
  {Akopian}, \citenamefont {Lindner}, \citenamefont {Poem}, \citenamefont
  {Berlatzky}, \citenamefont {Avron}, \citenamefont {Gershoni}, \citenamefont
  {Gerardot},\ and\ \citenamefont {Petroff}}]{Akopian2006}%
  \BibitemOpen
  \bibfield  {author} {\bibinfo {author} {\bibfnamefont {N.}~\bibnamefont
  {Akopian}}, \bibinfo {author} {\bibfnamefont {N.~H.}\ \bibnamefont
  {Lindner}}, \bibinfo {author} {\bibfnamefont {E.}~\bibnamefont {Poem}},
  \bibinfo {author} {\bibfnamefont {Y.}~\bibnamefont {Berlatzky}}, \bibinfo
  {author} {\bibfnamefont {J.}~\bibnamefont {Avron}}, \bibinfo {author}
  {\bibfnamefont {D.}~\bibnamefont {Gershoni}}, \bibinfo {author}
  {\bibfnamefont {B.~D.}\ \bibnamefont {Gerardot}}, \ and\ \bibinfo {author}
  {\bibfnamefont {P.~M.}\ \bibnamefont {Petroff}},\ }\href {\doibase
  10.1103/PhysRevLett.96.130501} {\bibfield  {journal} {\bibinfo  {journal}
  {Phys. Rev. Lett.}\ }\textbf {\bibinfo {volume} {96}},\ \bibinfo {pages}
  {130501} (\bibinfo {year} {2006})}\BibitemShut {NoStop}%
\bibitem [{\citenamefont {Stevenson}\ \emph
  {et~al.}(2006{\natexlab{a}})\citenamefont {Stevenson}, \citenamefont {Young},
  \citenamefont {Atkinson}, \citenamefont {Cooper}, \citenamefont {Ritchie},\
  and\ \citenamefont {Shields}}]{Stevenson2006}%
  \BibitemOpen
  \bibfield  {author} {\bibinfo {author} {\bibfnamefont {R.~M.}\ \bibnamefont
  {Stevenson}}, \bibinfo {author} {\bibfnamefont {R.~J.}\ \bibnamefont
  {Young}}, \bibinfo {author} {\bibfnamefont {P.}~\bibnamefont {Atkinson}},
  \bibinfo {author} {\bibfnamefont {K.}~\bibnamefont {Cooper}}, \bibinfo
  {author} {\bibfnamefont {D.~A.}\ \bibnamefont {Ritchie}}, \ and\ \bibinfo
  {author} {\bibfnamefont {A.~J.}\ \bibnamefont {Shields}},\ }\href
  {http://dx.doi.org/10.1038/nature04446} {\bibfield  {journal} {\bibinfo
  {journal} {Nature}\ }\textbf {\bibinfo {volume} {439}},\ \bibinfo {pages}
  {179} (\bibinfo {year} {2006}{\natexlab{a}})}\BibitemShut {NoStop}%
\bibitem [{\citenamefont {Hafenbrak}\ \emph {et~al.}(2007)\citenamefont
  {Hafenbrak}, \citenamefont {Ulrich}, \citenamefont {Michler}, \citenamefont
  {Wang}, \citenamefont {Rastelli},\ and\ \citenamefont
  {Schmidt}}]{Hafenbrak2007}%
  \BibitemOpen
  \bibfield  {author} {\bibinfo {author} {\bibfnamefont {R.}~\bibnamefont
  {Hafenbrak}}, \bibinfo {author} {\bibfnamefont {S.~M.}\ \bibnamefont
  {Ulrich}}, \bibinfo {author} {\bibfnamefont {P.}~\bibnamefont {Michler}},
  \bibinfo {author} {\bibfnamefont {L.}~\bibnamefont {Wang}}, \bibinfo {author}
  {\bibfnamefont {A.}~\bibnamefont {Rastelli}}, \ and\ \bibinfo {author}
  {\bibfnamefont {O.~G.}\ \bibnamefont {Schmidt}},\ }\href
  {http://stacks.iop.org/1367-2630/9/i=9/a=315} {\bibfield  {journal} {\bibinfo
   {journal} {New J. Phys.}\ }\textbf {\bibinfo {volume} {9}},\ \bibinfo
  {pages} {315} (\bibinfo {year} {2007})}\BibitemShut {NoStop}%
\bibitem [{\citenamefont {Dousse}\ \emph {et~al.}(2010)\citenamefont {Dousse},
  \citenamefont {Suffczynski}, \citenamefont {Beveratos}, \citenamefont
  {Krebs}, \citenamefont {Lema{\^i}tre}, \citenamefont {Sagnes}, \citenamefont
  {Bloch}, \citenamefont {Voisin},\ and\ \citenamefont
  {Senellart}}]{Dousse2010}%
  \BibitemOpen
  \bibfield  {author} {\bibinfo {author} {\bibfnamefont {A.}~\bibnamefont
  {Dousse}}, \bibinfo {author} {\bibfnamefont {J.}~\bibnamefont {Suffczynski}},
  \bibinfo {author} {\bibfnamefont {A.}~\bibnamefont {Beveratos}}, \bibinfo
  {author} {\bibfnamefont {O.}~\bibnamefont {Krebs}}, \bibinfo {author}
  {\bibfnamefont {A.}~\bibnamefont {Lema{\^i}tre}}, \bibinfo {author}
  {\bibfnamefont {I.}~\bibnamefont {Sagnes}}, \bibinfo {author} {\bibfnamefont
  {J.}~\bibnamefont {Bloch}}, \bibinfo {author} {\bibfnamefont
  {P.}~\bibnamefont {Voisin}}, \ and\ \bibinfo {author} {\bibfnamefont
  {P.}~\bibnamefont {Senellart}},\ }\href
  {http://dx.doi.org/10.1038/nature09148} {\bibfield  {journal} {\bibinfo
  {journal} {Nature}\ }\textbf {\bibinfo {volume} {466}},\ \bibinfo {pages}
  {217} (\bibinfo {year} {2010})}\BibitemShut {NoStop}%
\bibitem [{\citenamefont {del Valle}(2013)}]{delvalle2013dis}%
  \BibitemOpen
  \bibfield  {author} {\bibinfo {author} {\bibfnamefont {E.}~\bibnamefont {del
  Valle}},\ }\href {https://doi.org/10.1088%2F1367-2630%2F15%2F2%2F025019}
  {\bibfield  {journal} {\bibinfo  {journal} {New J.\ Phys}\ }\textbf {\bibinfo
  {volume} {15}},\ \bibinfo {pages} {025019} (\bibinfo {year}
  {2013})}\BibitemShut {NoStop}%
\bibitem [{\citenamefont {M{\"u}ller}\ \emph {et~al.}(2014)\citenamefont
  {M{\"u}ller}, \citenamefont {Bounouar}, \citenamefont {J{\"o}ns},
  \citenamefont {Gl{\"a}ssl},\ and\ \citenamefont {Michler}}]{Mueller2014}%
  \BibitemOpen
  \bibfield  {author} {\bibinfo {author} {\bibfnamefont {M.}~\bibnamefont
  {M{\"u}ller}}, \bibinfo {author} {\bibfnamefont {S.}~\bibnamefont
  {Bounouar}}, \bibinfo {author} {\bibfnamefont {K.~D.}\ \bibnamefont
  {J{\"o}ns}}, \bibinfo {author} {\bibfnamefont {M.}~\bibnamefont
  {Gl{\"a}ssl}}, \ and\ \bibinfo {author} {\bibfnamefont {P.}~\bibnamefont
  {Michler}},\ }\href {http://dx.doi.org/10.1038/nphoton.2013.377} {\bibfield
  {journal} {\bibinfo  {journal} {Nat. Photonics}\ }\textbf {\bibinfo {volume}
  {8}},\ \bibinfo {pages} {224} (\bibinfo {year} {2014})}\BibitemShut {NoStop}%
\bibitem [{\citenamefont {Orieux}\ \emph {et~al.}(2017)\citenamefont {Orieux},
  \citenamefont {Versteegh}, \citenamefont {J\"ons},\ and\ \citenamefont
  {Ducci}}]{Orieux2017}%
  \BibitemOpen
  \bibfield  {author} {\bibinfo {author} {\bibfnamefont {A.}~\bibnamefont
  {Orieux}}, \bibinfo {author} {\bibfnamefont {M.~A.~M.}\ \bibnamefont
  {Versteegh}}, \bibinfo {author} {\bibfnamefont {K.~D.}\ \bibnamefont
  {J\"ons}}, \ and\ \bibinfo {author} {\bibfnamefont {S.}~\bibnamefont
  {Ducci}},\ }\href {\doibase 10.1088/1361-6633/aa6955} {\bibfield  {journal}
  {\bibinfo  {journal} {Rep. Prog. Phys.}\ }\textbf {\bibinfo {volume} {80}},\
  \bibinfo {pages} {076001} (\bibinfo {year} {2017})}\BibitemShut {NoStop}%
\bibitem [{\citenamefont {Seidelmann}\ \emph {et~al.}(2019)\citenamefont
  {Seidelmann}, \citenamefont {Ungar}, \citenamefont {Barth}, \citenamefont
  {Vagov}, \citenamefont {Axt}, \citenamefont {Cygorek},\ and\ \citenamefont
  {Kuhn}}]{Seidelmann2019}%
  \BibitemOpen
  \bibfield  {author} {\bibinfo {author} {\bibfnamefont {T.}~\bibnamefont
  {Seidelmann}}, \bibinfo {author} {\bibfnamefont {F.}~\bibnamefont {Ungar}},
  \bibinfo {author} {\bibfnamefont {A.~M.}\ \bibnamefont {Barth}}, \bibinfo
  {author} {\bibfnamefont {A.}~\bibnamefont {Vagov}}, \bibinfo {author}
  {\bibfnamefont {V.~M.}\ \bibnamefont {Axt}}, \bibinfo {author} {\bibfnamefont
  {M.}~\bibnamefont {Cygorek}}, \ and\ \bibinfo {author} {\bibfnamefont
  {T.}~\bibnamefont {Kuhn}},\ }\href {\doibase 10.1103/PhysRevLett.123.137401}
  {\bibfield  {journal} {\bibinfo  {journal} {Phys. Rev. Lett.}\ }\textbf
  {\bibinfo {volume} {123}},\ \bibinfo {pages} {137401} (\bibinfo {year}
  {2019})}\BibitemShut {NoStop}%
\bibitem [{\citenamefont {Holland}\ and\ \citenamefont
  {Burnett}(1993)}]{Holland1993}%
  \BibitemOpen
  \bibfield  {author} {\bibinfo {author} {\bibfnamefont {M.~J.}\ \bibnamefont
  {Holland}}\ and\ \bibinfo {author} {\bibfnamefont {K.}~\bibnamefont
  {Burnett}},\ }\href {\doibase 10.1103/PhysRevLett.71.1355} {\bibfield
  {journal} {\bibinfo  {journal} {Phys. Rev. Lett.}\ }\textbf {\bibinfo
  {volume} {71}},\ \bibinfo {pages} {1355} (\bibinfo {year}
  {1993})}\BibitemShut {NoStop}%
\bibitem [{\citenamefont {Nagata}\ \emph {et~al.}(2007)\citenamefont {Nagata},
  \citenamefont {Okamoto}, \citenamefont {O'Brien}, \citenamefont {Sasaki},\
  and\ \citenamefont {Takeuchi}}]{Nagata2007}%
  \BibitemOpen
  \bibfield  {author} {\bibinfo {author} {\bibfnamefont {T.}~\bibnamefont
  {Nagata}}, \bibinfo {author} {\bibfnamefont {R.}~\bibnamefont {Okamoto}},
  \bibinfo {author} {\bibfnamefont {J.~L.}\ \bibnamefont {O'Brien}}, \bibinfo
  {author} {\bibfnamefont {K.}~\bibnamefont {Sasaki}}, \ and\ \bibinfo {author}
  {\bibfnamefont {S.}~\bibnamefont {Takeuchi}},\ }\href {\doibase
  10.1126/science.1138007} {\bibfield  {journal} {\bibinfo  {journal}
  {Science}\ }\textbf {\bibinfo {volume} {316}},\ \bibinfo {pages} {726}
  (\bibinfo {year} {2007})}\BibitemShut {NoStop}%
\bibitem [{\citenamefont {Slussarenko}\ \emph {et~al.}(2017)\citenamefont
  {Slussarenko}, \citenamefont {Weston}, \citenamefont {Chrzanowski},
  \citenamefont {Shalm}, \citenamefont {Verma}, \citenamefont {Nam},\ and\
  \citenamefont {Pryde}}]{Slussarenko2017}%
  \BibitemOpen
  \bibfield  {author} {\bibinfo {author} {\bibfnamefont {S.}~\bibnamefont
  {Slussarenko}}, \bibinfo {author} {\bibfnamefont {M.~M.}\ \bibnamefont
  {Weston}}, \bibinfo {author} {\bibfnamefont {H.~M.}\ \bibnamefont
  {Chrzanowski}}, \bibinfo {author} {\bibfnamefont {L.~K.}\ \bibnamefont
  {Shalm}}, \bibinfo {author} {\bibfnamefont {V.~B.}\ \bibnamefont {Verma}},
  \bibinfo {author} {\bibfnamefont {S.~W.}\ \bibnamefont {Nam}}, \ and\
  \bibinfo {author} {\bibfnamefont {G.~J.}\ \bibnamefont {Pryde}},\ }\href
  {\doibase 10.1038/s41566-017-0011-5} {\bibfield  {journal} {\bibinfo
  {journal} {Nature Photon}\ }\textbf {\bibinfo {volume} {11}},\ \bibinfo
  {pages} {700} (\bibinfo {year} {2017})}\BibitemShut {NoStop}%
\bibitem [{\citenamefont {Ourjoumtsev}\ \emph {et~al.}(2007)\citenamefont
  {Ourjoumtsev}, \citenamefont {Jeong}, \citenamefont {Tualle-Brouri},\ and\
  \citenamefont {Grangier}}]{Ourjoumtsev2007}%
  \BibitemOpen
  \bibfield  {author} {\bibinfo {author} {\bibfnamefont {A.}~\bibnamefont
  {Ourjoumtsev}}, \bibinfo {author} {\bibfnamefont {H.}~\bibnamefont {Jeong}},
  \bibinfo {author} {\bibfnamefont {R.}~\bibnamefont {Tualle-Brouri}}, \ and\
  \bibinfo {author} {\bibfnamefont {P.}~\bibnamefont {Grangier}},\ }\href
  {\doibase 10.1038/nature06054} {\bibfield  {journal} {\bibinfo  {journal}
  {Nature}\ }\textbf {\bibinfo {volume} {448}},\ \bibinfo {pages} {784}
  (\bibinfo {year} {2007})}\BibitemShut {NoStop}%
\bibitem [{\citenamefont {Yamamoto}\ and\ \citenamefont
  {Haus}(1986)}]{Yamamoto1986}%
  \BibitemOpen
  \bibfield  {author} {\bibinfo {author} {\bibfnamefont {Y.}~\bibnamefont
  {Yamamoto}}\ and\ \bibinfo {author} {\bibfnamefont {H.~A.}\ \bibnamefont
  {Haus}},\ }\href {\doibase 10.1103/RevModPhys.58.1001} {\bibfield  {journal}
  {\bibinfo  {journal} {Rev. Mod. Phys.}\ }\textbf {\bibinfo {volume} {58}},\
  \bibinfo {pages} {1001} (\bibinfo {year} {1986})}\BibitemShut {NoStop}%
\bibitem [{\citenamefont {Cummings}\ and\ \citenamefont
  {Rajagopal}(1989)}]{Cummings1989}%
  \BibitemOpen
  \bibfield  {author} {\bibinfo {author} {\bibfnamefont {F.~W.}\ \bibnamefont
  {Cummings}}\ and\ \bibinfo {author} {\bibfnamefont {A.~K.}\ \bibnamefont
  {Rajagopal}},\ }\href {\doibase 10.1103/PhysRevA.39.3414} {\bibfield
  {journal} {\bibinfo  {journal} {Phys. Rev. A}\ }\textbf {\bibinfo {volume}
  {39}},\ \bibinfo {pages} {3414} (\bibinfo {year} {1989})}\BibitemShut
  {NoStop}%
\bibitem [{\citenamefont {Varcoe}\ \emph {et~al.}(2000)\citenamefont {Varcoe},
  \citenamefont {Brattke}, \citenamefont {Weidinger},\ and\ \citenamefont
  {Walther}}]{Varcoe2000}%
  \BibitemOpen
  \bibfield  {author} {\bibinfo {author} {\bibfnamefont {B.~T.~H.}\
  \bibnamefont {Varcoe}}, \bibinfo {author} {\bibfnamefont {S.}~\bibnamefont
  {Brattke}}, \bibinfo {author} {\bibfnamefont {M.}~\bibnamefont {Weidinger}},
  \ and\ \bibinfo {author} {\bibfnamefont {H.}~\bibnamefont {Walther}},\ }\href
  {\doibase 10.1038/35001526} {\bibfield  {journal} {\bibinfo  {journal}
  {Nature}\ }\textbf {\bibinfo {volume} {403}},\ \bibinfo {pages} {743}
  (\bibinfo {year} {2000})}\BibitemShut {NoStop}%
\bibitem [{\citenamefont {Zhou}\ \emph {et~al.}(2012)\citenamefont {Zhou},
  \citenamefont {Dotsenko}, \citenamefont {Peaudecerf}, \citenamefont
  {Rybarczyk}, \citenamefont {Sayrin}, \citenamefont {Gleyzes}, \citenamefont
  {Raimond}, \citenamefont {Brune},\ and\ \citenamefont
  {Haroche}}]{Haroche2012}%
  \BibitemOpen
  \bibfield  {author} {\bibinfo {author} {\bibfnamefont {X.}~\bibnamefont
  {Zhou}}, \bibinfo {author} {\bibfnamefont {I.}~\bibnamefont {Dotsenko}},
  \bibinfo {author} {\bibfnamefont {B.}~\bibnamefont {Peaudecerf}}, \bibinfo
  {author} {\bibfnamefont {T.}~\bibnamefont {Rybarczyk}}, \bibinfo {author}
  {\bibfnamefont {C.}~\bibnamefont {Sayrin}}, \bibinfo {author} {\bibfnamefont
  {S.}~\bibnamefont {Gleyzes}}, \bibinfo {author} {\bibfnamefont {J.~M.}\
  \bibnamefont {Raimond}}, \bibinfo {author} {\bibfnamefont {M.}~\bibnamefont
  {Brune}}, \ and\ \bibinfo {author} {\bibfnamefont {S.}~\bibnamefont
  {Haroche}},\ }\href {\doibase 10.1103/PhysRevLett.108.243602} {\bibfield
  {journal} {\bibinfo  {journal} {Phys. Rev. Lett.}\ }\textbf {\bibinfo
  {volume} {108}},\ \bibinfo {pages} {243602} (\bibinfo {year}
  {2012})}\BibitemShut {NoStop}%
\bibitem [{\citenamefont {Hofheinz}\ \emph {et~al.}(2008)\citenamefont
  {Hofheinz}, \citenamefont {Weig}, \citenamefont {Ansmann}, \citenamefont
  {Bialczak}, \citenamefont {Lucero}, \citenamefont {Neeley}, \citenamefont
  {O’Connell}, \citenamefont {Wang}, \citenamefont {Martinis},\ and\
  \citenamefont {Cleland}}]{Hofheinz2008}%
  \BibitemOpen
  \bibfield  {author} {\bibinfo {author} {\bibfnamefont {M.}~\bibnamefont
  {Hofheinz}}, \bibinfo {author} {\bibfnamefont {E.~M.}\ \bibnamefont {Weig}},
  \bibinfo {author} {\bibfnamefont {M.}~\bibnamefont {Ansmann}}, \bibinfo
  {author} {\bibfnamefont {R.~C.}\ \bibnamefont {Bialczak}}, \bibinfo {author}
  {\bibfnamefont {E.}~\bibnamefont {Lucero}}, \bibinfo {author} {\bibfnamefont
  {M.}~\bibnamefont {Neeley}}, \bibinfo {author} {\bibfnamefont {A.~D.}\
  \bibnamefont {O’Connell}}, \bibinfo {author} {\bibfnamefont
  {H.}~\bibnamefont {Wang}}, \bibinfo {author} {\bibfnamefont {J.~M.}\
  \bibnamefont {Martinis}}, \ and\ \bibinfo {author} {\bibfnamefont {A.~N.}\
  \bibnamefont {Cleland}},\ }\href {\doibase 10.1038/nature07136} {\bibfield
  {journal} {\bibinfo  {journal} {Nature}\ }\textbf {\bibinfo {volume} {454}},\
  \bibinfo {pages} {310} (\bibinfo {year} {2008})}\BibitemShut {NoStop}%
\bibitem [{\citenamefont {Tiedau}\ \emph {et~al.}(2019)\citenamefont {Tiedau},
  \citenamefont {Bartley}, \citenamefont {Harder}, \citenamefont {Lita},
  \citenamefont {Nam}, \citenamefont {Gerrits},\ and\ \citenamefont
  {Silberhorn}}]{Tiedau2019}%
  \BibitemOpen
  \bibfield  {author} {\bibinfo {author} {\bibfnamefont {J.}~\bibnamefont
  {Tiedau}}, \bibinfo {author} {\bibfnamefont {T.~J.}\ \bibnamefont {Bartley}},
  \bibinfo {author} {\bibfnamefont {G.}~\bibnamefont {Harder}}, \bibinfo
  {author} {\bibfnamefont {A.~E.}\ \bibnamefont {Lita}}, \bibinfo {author}
  {\bibfnamefont {S.~W.}\ \bibnamefont {Nam}}, \bibinfo {author} {\bibfnamefont
  {T.}~\bibnamefont {Gerrits}}, \ and\ \bibinfo {author} {\bibfnamefont
  {C.}~\bibnamefont {Silberhorn}},\ }\href {\doibase
  10.1103/PhysRevA.100.041802} {\bibfield  {journal} {\bibinfo  {journal}
  {Phys. Rev. A}\ }\textbf {\bibinfo {volume} {100}},\ \bibinfo {pages}
  {041802(R)} (\bibinfo {year} {2019})}\BibitemShut {NoStop}%
\bibitem [{\citenamefont {Besombes}\ \emph {et~al.}(2001)\citenamefont
  {Besombes}, \citenamefont {Kheng}, \citenamefont {Marsal},\ and\
  \citenamefont {Mariette}}]{Besombes2001}%
  \BibitemOpen
  \bibfield  {author} {\bibinfo {author} {\bibfnamefont {L.}~\bibnamefont
  {Besombes}}, \bibinfo {author} {\bibfnamefont {K.}~\bibnamefont {Kheng}},
  \bibinfo {author} {\bibfnamefont {L.}~\bibnamefont {Marsal}}, \ and\ \bibinfo
  {author} {\bibfnamefont {H.}~\bibnamefont {Mariette}},\ }\href {\doibase
  10.1103/PhysRevB.63.155307} {\bibfield  {journal} {\bibinfo  {journal} {Phys.
  Rev. B}\ }\textbf {\bibinfo {volume} {63}},\ \bibinfo {pages} {155307}
  (\bibinfo {year} {2001})}\BibitemShut {NoStop}%
\bibitem [{\citenamefont {Borri}\ \emph {et~al.}(2001)\citenamefont {Borri},
  \citenamefont {Langbein}, \citenamefont {Schneider}, \citenamefont {Woggon},
  \citenamefont {Sellin}, \citenamefont {Ouyang},\ and\ \citenamefont
  {Bimberg}}]{Borri2001}%
  \BibitemOpen
  \bibfield  {author} {\bibinfo {author} {\bibfnamefont {P.}~\bibnamefont
  {Borri}}, \bibinfo {author} {\bibfnamefont {W.}~\bibnamefont {Langbein}},
  \bibinfo {author} {\bibfnamefont {S.}~\bibnamefont {Schneider}}, \bibinfo
  {author} {\bibfnamefont {U.}~\bibnamefont {Woggon}}, \bibinfo {author}
  {\bibfnamefont {R.~L.}\ \bibnamefont {Sellin}}, \bibinfo {author}
  {\bibfnamefont {D.}~\bibnamefont {Ouyang}}, \ and\ \bibinfo {author}
  {\bibfnamefont {D.}~\bibnamefont {Bimberg}},\ }\href {\doibase
  10.1103/PhysRevLett.87.157401} {\bibfield  {journal} {\bibinfo  {journal}
  {Phys. Rev. Lett.}\ }\textbf {\bibinfo {volume} {87}},\ \bibinfo {pages}
  {157401} (\bibinfo {year} {2001})}\BibitemShut {NoStop}%
\bibitem [{\citenamefont {Krummheuer}\ \emph {et~al.}(2002)\citenamefont
  {Krummheuer}, \citenamefont {Axt},\ and\ \citenamefont
  {Kuhn}}]{Krummheuer2002}%
  \BibitemOpen
  \bibfield  {author} {\bibinfo {author} {\bibfnamefont {B.}~\bibnamefont
  {Krummheuer}}, \bibinfo {author} {\bibfnamefont {V.~M.}\ \bibnamefont {Axt}},
  \ and\ \bibinfo {author} {\bibfnamefont {T.}~\bibnamefont {Kuhn}},\ }\href
  {\doibase 10.1103/PhysRevB.65.195313} {\bibfield  {journal} {\bibinfo
  {journal} {Phys. Rev. B}\ }\textbf {\bibinfo {volume} {65}},\ \bibinfo
  {pages} {195313} (\bibinfo {year} {2002})}\BibitemShut {NoStop}%
\bibitem [{\citenamefont {Axt}\ \emph {et~al.}(2005)\citenamefont {Axt},
  \citenamefont {Kuhn}, \citenamefont {Vagov},\ and\ \citenamefont
  {Peeters}}]{Axt2005}%
  \BibitemOpen
  \bibfield  {author} {\bibinfo {author} {\bibfnamefont {V.~M.}\ \bibnamefont
  {Axt}}, \bibinfo {author} {\bibfnamefont {T.}~\bibnamefont {Kuhn}}, \bibinfo
  {author} {\bibfnamefont {A.}~\bibnamefont {Vagov}}, \ and\ \bibinfo {author}
  {\bibfnamefont {F.~M.}\ \bibnamefont {Peeters}},\ }\href {\doibase
  10.1103/PhysRevB.72.125309} {\bibfield  {journal} {\bibinfo  {journal} {Phys.
  Rev. B}\ }\textbf {\bibinfo {volume} {72}},\ \bibinfo {pages} {125309}
  (\bibinfo {year} {2005})}\BibitemShut {NoStop}%
\bibitem [{\citenamefont {Reiter}\ \emph {et~al.}(2019)\citenamefont {Reiter},
  \citenamefont {Kuhn},\ and\ \citenamefont {Axt}}]{Reiter2019}%
  \BibitemOpen
  \bibfield  {author} {\bibinfo {author} {\bibfnamefont {D.~E.}\ \bibnamefont
  {Reiter}}, \bibinfo {author} {\bibfnamefont {T.}~\bibnamefont {Kuhn}}, \ and\
  \bibinfo {author} {\bibfnamefont {V.~M.}\ \bibnamefont {Axt}},\ }\href
  {https://doi.org/10.1080/23746149.2019.1655478} {\bibfield  {journal}
  {\bibinfo  {journal} {Advances in Physics: X}\ }\textbf {\bibinfo {volume}
  {4}},\ \bibinfo {pages} {1655478} (\bibinfo {year} {2019})}\BibitemShut
  {NoStop}%
\bibitem [{\citenamefont {Stevenson}\ \emph
  {et~al.}(2006{\natexlab{b}})\citenamefont {Stevenson}, \citenamefont {Young},
  \citenamefont {Atkinson}, \citenamefont {Cooper}, \citenamefont {Ritchie},\
  and\ \citenamefont {Shields}}]{Stevenson2006a}%
  \BibitemOpen
  \bibfield  {author} {\bibinfo {author} {\bibfnamefont {R.~M.}\ \bibnamefont
  {Stevenson}}, \bibinfo {author} {\bibfnamefont {R.~J.}\ \bibnamefont
  {Young}}, \bibinfo {author} {\bibfnamefont {P.}~\bibnamefont {Atkinson}},
  \bibinfo {author} {\bibfnamefont {K.}~\bibnamefont {Cooper}}, \bibinfo
  {author} {\bibfnamefont {D.~A.}\ \bibnamefont {Ritchie}}, \ and\ \bibinfo
  {author} {\bibfnamefont {A.~J.}\ \bibnamefont {Shields}},\ }\href {\doibase
  10.1038/nature04446} {\bibfield  {journal} {\bibinfo  {journal} {Nature}\
  }\textbf {\bibinfo {volume} {439}},\ \bibinfo {pages} {179} (\bibinfo {year}
  {2006}{\natexlab{b}})}\BibitemShut {NoStop}%
\bibitem [{\citenamefont {Young}\ \emph {et~al.}(2006)\citenamefont {Young},
  \citenamefont {Stevenson}, \citenamefont {Atkinson}, \citenamefont {Cooper},
  \citenamefont {Ritchie},\ and\ \citenamefont {Shields}}]{Young2006}%
  \BibitemOpen
  \bibfield  {author} {\bibinfo {author} {\bibfnamefont {R.~J.}\ \bibnamefont
  {Young}}, \bibinfo {author} {\bibfnamefont {R.~M.}\ \bibnamefont
  {Stevenson}}, \bibinfo {author} {\bibfnamefont {P.}~\bibnamefont {Atkinson}},
  \bibinfo {author} {\bibfnamefont {K.}~\bibnamefont {Cooper}}, \bibinfo
  {author} {\bibfnamefont {D.~A.}\ \bibnamefont {Ritchie}}, \ and\ \bibinfo
  {author} {\bibfnamefont {A.~J.}\ \bibnamefont {Shields}},\ }\href {\doibase
  10.1088/1367-2630/8/2/029} {\bibfield  {journal} {\bibinfo  {journal} {New J.
  Phys.}\ }\textbf {\bibinfo {volume} {8}},\ \bibinfo {pages} {29} (\bibinfo
  {year} {2006})}\BibitemShut {NoStop}%
\bibitem [{\citenamefont {Stevenson}\ \emph
  {et~al.}(2006{\natexlab{c}})\citenamefont {Stevenson}, \citenamefont {Young},
  \citenamefont {See}, \citenamefont {Gevaux}, \citenamefont {Cooper},
  \citenamefont {Atkinson}, \citenamefont {Farrer}, \citenamefont {Ritchie},\
  and\ \citenamefont {Shields}}]{Stevenson2006b}%
  \BibitemOpen
  \bibfield  {author} {\bibinfo {author} {\bibfnamefont {R.~M.}\ \bibnamefont
  {Stevenson}}, \bibinfo {author} {\bibfnamefont {R.~J.}\ \bibnamefont
  {Young}}, \bibinfo {author} {\bibfnamefont {P.}~\bibnamefont {See}}, \bibinfo
  {author} {\bibfnamefont {D.~G.}\ \bibnamefont {Gevaux}}, \bibinfo {author}
  {\bibfnamefont {K.}~\bibnamefont {Cooper}}, \bibinfo {author} {\bibfnamefont
  {P.}~\bibnamefont {Atkinson}}, \bibinfo {author} {\bibfnamefont
  {I.}~\bibnamefont {Farrer}}, \bibinfo {author} {\bibfnamefont {D.~A.}\
  \bibnamefont {Ritchie}}, \ and\ \bibinfo {author} {\bibfnamefont {A.~J.}\
  \bibnamefont {Shields}},\ }\href {\doibase 10.1103/PhysRevB.73.033306}
  {\bibfield  {journal} {\bibinfo  {journal} {Phys. Rev. B}\ }\textbf {\bibinfo
  {volume} {73}},\ \bibinfo {pages} {033306} (\bibinfo {year}
  {2006}{\natexlab{c}})}\BibitemShut {NoStop}%
\bibitem [{\citenamefont {Zhang}\ \emph {et~al.}(2015)\citenamefont {Zhang},
  \citenamefont {Wildmann}, \citenamefont {Ding}, \citenamefont {Trotta},
  \citenamefont {Huo}, \citenamefont {Zallo}, \citenamefont {Huber},
  \citenamefont {Rastelli},\ and\ \citenamefont {Schmidt}}]{Zhang2015}%
  \BibitemOpen
  \bibfield  {author} {\bibinfo {author} {\bibfnamefont {J.}~\bibnamefont
  {Zhang}}, \bibinfo {author} {\bibfnamefont {J.~S.}\ \bibnamefont {Wildmann}},
  \bibinfo {author} {\bibfnamefont {F.}~\bibnamefont {Ding}}, \bibinfo {author}
  {\bibfnamefont {R.}~\bibnamefont {Trotta}}, \bibinfo {author} {\bibfnamefont
  {Y.}~\bibnamefont {Huo}}, \bibinfo {author} {\bibfnamefont {E.}~\bibnamefont
  {Zallo}}, \bibinfo {author} {\bibfnamefont {D.}~\bibnamefont {Huber}},
  \bibinfo {author} {\bibfnamefont {A.}~\bibnamefont {Rastelli}}, \ and\
  \bibinfo {author} {\bibfnamefont {O.~G.}\ \bibnamefont {Schmidt}},\ }\href
  {\doibase 10.1038/ncomms10067} {\bibfield  {journal} {\bibinfo  {journal}
  {Nat Commun}\ }\textbf {\bibinfo {volume} {6}},\ \bibinfo {pages} {10067}
  (\bibinfo {year} {2015})}\BibitemShut {NoStop}%
\bibitem [{\citenamefont {Cygorek}\ \emph {et~al.}(2020)\citenamefont
  {Cygorek}, \citenamefont {Korkusinski},\ and\ \citenamefont
  {Hawrylak}}]{Cygorek2020}%
  \BibitemOpen
  \bibfield  {author} {\bibinfo {author} {\bibfnamefont {M.}~\bibnamefont
  {Cygorek}}, \bibinfo {author} {\bibfnamefont {M.}~\bibnamefont
  {Korkusinski}}, \ and\ \bibinfo {author} {\bibfnamefont {P.}~\bibnamefont
  {Hawrylak}},\ }\href {\doibase 10.1103/PhysRevB.101.075307} {\bibfield
  {journal} {\bibinfo  {journal} {Phys. Rev. B}\ }\textbf {\bibinfo {volume}
  {101}},\ \bibinfo {pages} {075307} (\bibinfo {year} {2020})}\BibitemShut
  {NoStop}%
\bibitem [{\citenamefont {Mukherjee}\ \emph {et~al.}(2020)\citenamefont
  {Mukherjee}, \citenamefont {Widhalm}, \citenamefont {Siebert}, \citenamefont
  {Krehs}, \citenamefont {Sharma}, \citenamefont {Thiede}, \citenamefont
  {Reuter}, \citenamefont {Förstner},\ and\ \citenamefont
  {Zrenner}}]{Mukherjee2020}%
  \BibitemOpen
  \bibfield  {author} {\bibinfo {author} {\bibfnamefont {A.}~\bibnamefont
  {Mukherjee}}, \bibinfo {author} {\bibfnamefont {A.}~\bibnamefont {Widhalm}},
  \bibinfo {author} {\bibfnamefont {D.}~\bibnamefont {Siebert}}, \bibinfo
  {author} {\bibfnamefont {S.}~\bibnamefont {Krehs}}, \bibinfo {author}
  {\bibfnamefont {N.}~\bibnamefont {Sharma}}, \bibinfo {author} {\bibfnamefont
  {A.}~\bibnamefont {Thiede}}, \bibinfo {author} {\bibfnamefont
  {D.}~\bibnamefont {Reuter}}, \bibinfo {author} {\bibfnamefont
  {J.}~\bibnamefont {Förstner}}, \ and\ \bibinfo {author} {\bibfnamefont
  {A.}~\bibnamefont {Zrenner}},\ }\href {\doibase 10.1063/5.0012257} {\bibfield
   {journal} {\bibinfo  {journal} {Applied Physics Letters}\ }\textbf {\bibinfo
  {volume} {116}},\ \bibinfo {pages} {251103} (\bibinfo {year}
  {2020})}\BibitemShut {NoStop}%
\bibitem [{\citenamefont {Vagov}\ \emph {et~al.}(2011)\citenamefont {Vagov},
  \citenamefont {Croitoru}, \citenamefont {Gl\"assl}, \citenamefont {Axt},\
  and\ \citenamefont {Kuhn}}]{Vagov2011}%
  \BibitemOpen
  \bibfield  {author} {\bibinfo {author} {\bibfnamefont {A.}~\bibnamefont
  {Vagov}}, \bibinfo {author} {\bibfnamefont {M.~D.}\ \bibnamefont {Croitoru}},
  \bibinfo {author} {\bibfnamefont {M.}~\bibnamefont {Gl\"assl}}, \bibinfo
  {author} {\bibfnamefont {V.~M.}\ \bibnamefont {Axt}}, \ and\ \bibinfo
  {author} {\bibfnamefont {T.}~\bibnamefont {Kuhn}},\ }\href {\doibase
  10.1103/PhysRevB.83.094303} {\bibfield  {journal} {\bibinfo  {journal} {Phys.
  Rev. B}\ }\textbf {\bibinfo {volume} {83}},\ \bibinfo {pages} {094303}
  (\bibinfo {year} {2011})}\BibitemShut {NoStop}%
\bibitem [{\citenamefont {Barth}\ \emph
  {et~al.}(2016{\natexlab{a}})\citenamefont {Barth}, \citenamefont {Vagov},\
  and\ \citenamefont {Axt}}]{Barth2016}%
  \BibitemOpen
  \bibfield  {author} {\bibinfo {author} {\bibfnamefont {A.~M.}\ \bibnamefont
  {Barth}}, \bibinfo {author} {\bibfnamefont {A.}~\bibnamefont {Vagov}}, \ and\
  \bibinfo {author} {\bibfnamefont {V.~M.}\ \bibnamefont {Axt}},\ }\href
  {\doibase 10.1103/PhysRevB.94.125439} {\bibfield  {journal} {\bibinfo
  {journal} {Phys. Rev. B}\ }\textbf {\bibinfo {volume} {94}},\ \bibinfo
  {pages} {125439} (\bibinfo {year} {2016}{\natexlab{a}})}\BibitemShut
  {NoStop}%
\bibitem [{\citenamefont {Cygorek}\ \emph {et~al.}(2017)\citenamefont
  {Cygorek}, \citenamefont {Barth}, \citenamefont {Ungar}, \citenamefont
  {Vagov},\ and\ \citenamefont {Axt}}]{Cygorek2017}%
  \BibitemOpen
  \bibfield  {author} {\bibinfo {author} {\bibfnamefont {M.}~\bibnamefont
  {Cygorek}}, \bibinfo {author} {\bibfnamefont {A.~M.}\ \bibnamefont {Barth}},
  \bibinfo {author} {\bibfnamefont {F.}~\bibnamefont {Ungar}}, \bibinfo
  {author} {\bibfnamefont {A.}~\bibnamefont {Vagov}}, \ and\ \bibinfo {author}
  {\bibfnamefont {V.~M.}\ \bibnamefont {Axt}},\ }\href {\doibase
  10.1103/PhysRevB.96.201201} {\bibfield  {journal} {\bibinfo  {journal} {Phys.
  Rev. B}\ }\textbf {\bibinfo {volume} {96}},\ \bibinfo {pages} {201201(R)}
  (\bibinfo {year} {2017})}\BibitemShut {NoStop}%
\bibitem [{\citenamefont {Schneider}\ \emph {et~al.}(2009)\citenamefont
  {Schneider}, \citenamefont {Heindel}, \citenamefont {Huggenberger},
  \citenamefont {Weinmann}, \citenamefont {Kistner}, \citenamefont {Kamp},
  \citenamefont {Reitzenstein}, \citenamefont {H{\"o}fling},\ and\
  \citenamefont {Forchel}}]{schneider2009single}%
  \BibitemOpen
  \bibfield  {author} {\bibinfo {author} {\bibfnamefont {C.}~\bibnamefont
  {Schneider}}, \bibinfo {author} {\bibfnamefont {T.}~\bibnamefont {Heindel}},
  \bibinfo {author} {\bibfnamefont {A.}~\bibnamefont {Huggenberger}}, \bibinfo
  {author} {\bibfnamefont {P.}~\bibnamefont {Weinmann}}, \bibinfo {author}
  {\bibfnamefont {C.}~\bibnamefont {Kistner}}, \bibinfo {author} {\bibfnamefont
  {M.}~\bibnamefont {Kamp}}, \bibinfo {author} {\bibfnamefont {S.}~\bibnamefont
  {Reitzenstein}}, \bibinfo {author} {\bibfnamefont {S.}~\bibnamefont
  {H{\"o}fling}}, \ and\ \bibinfo {author} {\bibfnamefont {A.}~\bibnamefont
  {Forchel}},\ }\href@noop {} {\bibfield  {journal} {\bibinfo  {journal}
  {Appl.\ Phys.\ Lett.}\ }\textbf {\bibinfo {volume} {94}},\ \bibinfo {pages}
  {111111} (\bibinfo {year} {2009})}\BibitemShut {NoStop}%
\bibitem [{\citenamefont {Reitzenstein}\ and\ \citenamefont
  {Forchel}(2010)}]{reitzenstein2010quantum}%
  \BibitemOpen
  \bibfield  {author} {\bibinfo {author} {\bibfnamefont {S.}~\bibnamefont
  {Reitzenstein}}\ and\ \bibinfo {author} {\bibfnamefont {A.}~\bibnamefont
  {Forchel}},\ }\href@noop {} {\ \textbf {\bibinfo {volume} {43}},\ \bibinfo
  {pages} {033001} (\bibinfo {year} {2010})}\BibitemShut {NoStop}%
\bibitem [{\citenamefont {Jozsa}(1994)}]{Jozsa1994}%
  \BibitemOpen
  \bibfield  {author} {\bibinfo {author} {\bibfnamefont {R.}~\bibnamefont
  {Jozsa}},\ }\href {\doibase 10.1080/09500349414552171} {\bibfield  {journal}
  {\bibinfo  {journal} {Journal of Modern Optics}\ }\textbf {\bibinfo {volume}
  {41}},\ \bibinfo {pages} {2315} (\bibinfo {year} {1994})}\BibitemShut
  {NoStop}%
\bibitem [{\citenamefont {Kaldewey}\ \emph {et~al.}(2017)\citenamefont
  {Kaldewey}, \citenamefont {L{\"u}ker}, \citenamefont {Kuhlmann},
  \citenamefont {Valentin}, \citenamefont {Chauveau}, \citenamefont {Ludwig},
  \citenamefont {Wieck}, \citenamefont {Reiter}, \citenamefont {Kuhn},\ and\
  \citenamefont {Warburton}}]{kaldewey2017demonstrating}%
  \BibitemOpen
  \bibfield  {author} {\bibinfo {author} {\bibfnamefont {T.}~\bibnamefont
  {Kaldewey}}, \bibinfo {author} {\bibfnamefont {S.}~\bibnamefont {L{\"u}ker}},
  \bibinfo {author} {\bibfnamefont {A.~V.}\ \bibnamefont {Kuhlmann}}, \bibinfo
  {author} {\bibfnamefont {S.~R.}\ \bibnamefont {Valentin}}, \bibinfo {author}
  {\bibfnamefont {J.-M.}\ \bibnamefont {Chauveau}}, \bibinfo {author}
  {\bibfnamefont {A.}~\bibnamefont {Ludwig}}, \bibinfo {author} {\bibfnamefont
  {A.~D.}\ \bibnamefont {Wieck}}, \bibinfo {author} {\bibfnamefont {D.~E.}\
  \bibnamefont {Reiter}}, \bibinfo {author} {\bibfnamefont {T.}~\bibnamefont
  {Kuhn}}, \ and\ \bibinfo {author} {\bibfnamefont {R.~J.}\ \bibnamefont
  {Warburton}},\ }\href@noop {} {\bibfield  {journal} {\bibinfo  {journal}
  {Phys.\ Rev.\ {\rm B}}\ }\textbf {\bibinfo {volume} {95}},\ \bibinfo {pages}
  {241306(R)} (\bibinfo {year} {2017})}\BibitemShut {NoStop}%
\bibitem [{\citenamefont {Bayer}\ \emph {et~al.}(1999)\citenamefont {Bayer},
  \citenamefont {Kuther}, \citenamefont {Forchel}, \citenamefont {Gorbunov},
  \citenamefont {Timofeev}, \citenamefont {Sch\"afer}, \citenamefont
  {Reithmaier}, \citenamefont {Reinecke},\ and\ \citenamefont
  {Walck}}]{Bayer1999}%
  \BibitemOpen
  \bibfield  {author} {\bibinfo {author} {\bibfnamefont {M.}~\bibnamefont
  {Bayer}}, \bibinfo {author} {\bibfnamefont {A.}~\bibnamefont {Kuther}},
  \bibinfo {author} {\bibfnamefont {A.}~\bibnamefont {Forchel}}, \bibinfo
  {author} {\bibfnamefont {A.}~\bibnamefont {Gorbunov}}, \bibinfo {author}
  {\bibfnamefont {V.~B.}\ \bibnamefont {Timofeev}}, \bibinfo {author}
  {\bibfnamefont {F.}~\bibnamefont {Sch\"afer}}, \bibinfo {author}
  {\bibfnamefont {J.~P.}\ \bibnamefont {Reithmaier}}, \bibinfo {author}
  {\bibfnamefont {T.~L.}\ \bibnamefont {Reinecke}}, \ and\ \bibinfo {author}
  {\bibfnamefont {S.~N.}\ \bibnamefont {Walck}},\ }\href {\doibase
  10.1103/PhysRevLett.82.1748} {\bibfield  {journal} {\bibinfo  {journal}
  {Phys. Rev. Lett.}\ }\textbf {\bibinfo {volume} {82}},\ \bibinfo {pages}
  {1748} (\bibinfo {year} {1999})}\BibitemShut {NoStop}%
\bibitem [{\citenamefont {H\"ogele}\ \emph {et~al.}(2004)\citenamefont
  {H\"ogele}, \citenamefont {Seidl}, \citenamefont {Kroner}, \citenamefont
  {Karrai}, \citenamefont {Warburton}, \citenamefont {Gerardot},\ and\
  \citenamefont {Petroff}}]{Hoegele2004}%
  \BibitemOpen
  \bibfield  {author} {\bibinfo {author} {\bibfnamefont {A.}~\bibnamefont
  {H\"ogele}}, \bibinfo {author} {\bibfnamefont {S.}~\bibnamefont {Seidl}},
  \bibinfo {author} {\bibfnamefont {M.}~\bibnamefont {Kroner}}, \bibinfo
  {author} {\bibfnamefont {K.}~\bibnamefont {Karrai}}, \bibinfo {author}
  {\bibfnamefont {R.~J.}\ \bibnamefont {Warburton}}, \bibinfo {author}
  {\bibfnamefont {B.~D.}\ \bibnamefont {Gerardot}}, \ and\ \bibinfo {author}
  {\bibfnamefont {P.~M.}\ \bibnamefont {Petroff}},\ }\href {\doibase
  10.1103/PhysRevLett.93.217401} {\bibfield  {journal} {\bibinfo  {journal}
  {Phys. Rev. Lett.}\ }\textbf {\bibinfo {volume} {93}},\ \bibinfo {pages}
  {217401} (\bibinfo {year} {2004})}\BibitemShut {NoStop}%
\bibitem [{\citenamefont {Patton}\ \emph {et~al.}(2003)\citenamefont {Patton},
  \citenamefont {Langbein},\ and\ \citenamefont {Woggon}}]{Patton2003}%
  \BibitemOpen
  \bibfield  {author} {\bibinfo {author} {\bibfnamefont {B.}~\bibnamefont
  {Patton}}, \bibinfo {author} {\bibfnamefont {W.}~\bibnamefont {Langbein}}, \
  and\ \bibinfo {author} {\bibfnamefont {U.}~\bibnamefont {Woggon}},\ }\href
  {\doibase 10.1103/PhysRevB.68.125316} {\bibfield  {journal} {\bibinfo
  {journal} {Phys. Rev. B}\ }\textbf {\bibinfo {volume} {68}},\ \bibinfo
  {pages} {125316} (\bibinfo {year} {2003})}\BibitemShut {NoStop}%
\bibitem [{\citenamefont {Gammon}\ \emph {et~al.}(1996)\citenamefont {Gammon},
  \citenamefont {Snow}, \citenamefont {Shanabrook}, \citenamefont {Katzer},\
  and\ \citenamefont {Park}}]{Gammon1996}%
  \BibitemOpen
  \bibfield  {author} {\bibinfo {author} {\bibfnamefont {D.}~\bibnamefont
  {Gammon}}, \bibinfo {author} {\bibfnamefont {E.~S.}\ \bibnamefont {Snow}},
  \bibinfo {author} {\bibfnamefont {B.~V.}\ \bibnamefont {Shanabrook}},
  \bibinfo {author} {\bibfnamefont {D.~S.}\ \bibnamefont {Katzer}}, \ and\
  \bibinfo {author} {\bibfnamefont {D.}~\bibnamefont {Park}},\ }\href {\doibase
  10.1103/PhysRevLett.76.3005} {\bibfield  {journal} {\bibinfo  {journal}
  {Phys. Rev. Lett.}\ }\textbf {\bibinfo {volume} {76}},\ \bibinfo {pages}
  {3005} (\bibinfo {year} {1996})}\BibitemShut {NoStop}%
\bibitem [{\citenamefont {Vagov}\ \emph {et~al.}(2004)\citenamefont {Vagov},
  \citenamefont {Axt}, \citenamefont {Kuhn}, \citenamefont {Langbein},
  \citenamefont {Borri},\ and\ \citenamefont {Woggon}}]{Vagov2004}%
  \BibitemOpen
  \bibfield  {author} {\bibinfo {author} {\bibfnamefont {A.}~\bibnamefont
  {Vagov}}, \bibinfo {author} {\bibfnamefont {V.~M.}\ \bibnamefont {Axt}},
  \bibinfo {author} {\bibfnamefont {T.}~\bibnamefont {Kuhn}}, \bibinfo {author}
  {\bibfnamefont {W.}~\bibnamefont {Langbein}}, \bibinfo {author}
  {\bibfnamefont {P.}~\bibnamefont {Borri}}, \ and\ \bibinfo {author}
  {\bibfnamefont {U.}~\bibnamefont {Woggon}},\ }\href {\doibase
  10.1103/PhysRevB.70.201305} {\bibfield  {journal} {\bibinfo  {journal} {Phys.
  Rev. B}\ }\textbf {\bibinfo {volume} {70}},\ \bibinfo {pages} {201305(R)}
  (\bibinfo {year} {2004})}\BibitemShut {NoStop}%
\bibitem [{\citenamefont {Gustin}\ and\ \citenamefont
  {Hughes}(2018)}]{Gustin2018}%
  \BibitemOpen
  \bibfield  {author} {\bibinfo {author} {\bibfnamefont {C.}~\bibnamefont
  {Gustin}}\ and\ \bibinfo {author} {\bibfnamefont {S.}~\bibnamefont
  {Hughes}},\ }\href {\doibase 10.1103/PhysRevB.98.045309} {\bibfield
  {journal} {\bibinfo  {journal} {Phys. Rev. B}\ }\textbf {\bibinfo {volume}
  {98}},\ \bibinfo {pages} {045309} (\bibinfo {year} {2018})}\BibitemShut
  {NoStop}%
\bibitem [{\citenamefont {Gawarecki}\ \emph {et~al.}(2012)\citenamefont
  {Gawarecki}, \citenamefont {L\"uker}, \citenamefont {Reiter}, \citenamefont
  {Kuhn}, \citenamefont {Gl\"assl}, \citenamefont {Axt}, \citenamefont
  {Grodecka-Grad},\ and\ \citenamefont {Machnikowski}}]{Gawarecki2012}%
  \BibitemOpen
  \bibfield  {author} {\bibinfo {author} {\bibfnamefont {K.}~\bibnamefont
  {Gawarecki}}, \bibinfo {author} {\bibfnamefont {S.}~\bibnamefont {L\"uker}},
  \bibinfo {author} {\bibfnamefont {D.~E.}\ \bibnamefont {Reiter}}, \bibinfo
  {author} {\bibfnamefont {T.}~\bibnamefont {Kuhn}}, \bibinfo {author}
  {\bibfnamefont {M.}~\bibnamefont {Gl\"assl}}, \bibinfo {author}
  {\bibfnamefont {V.~M.}\ \bibnamefont {Axt}}, \bibinfo {author} {\bibfnamefont
  {A.}~\bibnamefont {Grodecka-Grad}}, \ and\ \bibinfo {author} {\bibfnamefont
  {P.}~\bibnamefont {Machnikowski}},\ }\href {\doibase
  10.1103/PhysRevB.86.235301} {\bibfield  {journal} {\bibinfo  {journal} {Phys.
  Rev. B}\ }\textbf {\bibinfo {volume} {86}},\ \bibinfo {pages} {235301}
  (\bibinfo {year} {2012})}\BibitemShut {NoStop}%
\bibitem [{\citenamefont {Chen}\ \emph {et~al.}(2001)\citenamefont {Chen},
  \citenamefont {Piermarocchi},\ and\ \citenamefont {Sham}}]{Chen2001}%
  \BibitemOpen
  \bibfield  {author} {\bibinfo {author} {\bibfnamefont {P.}~\bibnamefont
  {Chen}}, \bibinfo {author} {\bibfnamefont {C.}~\bibnamefont {Piermarocchi}},
  \ and\ \bibinfo {author} {\bibfnamefont {L.~J.}\ \bibnamefont {Sham}},\
  }\href {\doibase 10.1103/PhysRevLett.87.067401} {\bibfield  {journal}
  {\bibinfo  {journal} {Phys. Rev. Lett.}\ }\textbf {\bibinfo {volume} {87}},\
  \bibinfo {pages} {067401} (\bibinfo {year} {2001})}\BibitemShut {NoStop}%
\bibitem [{\citenamefont {{Dialynas}}\ \emph {et~al.}(2007)\citenamefont
  {{Dialynas}}, \citenamefont {{Xenogianni}}, \citenamefont {{Trichas}},
  \citenamefont {{Savvidis}}, \citenamefont {{Constantinidis}}, \citenamefont
  {{Hatzopoulos}},\ and\ \citenamefont {{Pelekanos}}}]{Dialynas2007}%
  \BibitemOpen
  \bibfield  {author} {\bibinfo {author} {\bibfnamefont {G.~E.}\ \bibnamefont
  {{Dialynas}}}, \bibinfo {author} {\bibfnamefont {C.}~\bibnamefont
  {{Xenogianni}}}, \bibinfo {author} {\bibfnamefont {E.}~\bibnamefont
  {{Trichas}}}, \bibinfo {author} {\bibfnamefont {P.~G.}\ \bibnamefont
  {{Savvidis}}}, \bibinfo {author} {\bibfnamefont {G.}~\bibnamefont
  {{Constantinidis}}}, \bibinfo {author} {\bibfnamefont {Z.}~\bibnamefont
  {{Hatzopoulos}}}, \ and\ \bibinfo {author} {\bibfnamefont {N.~T.}\
  \bibnamefont {{Pelekanos}}},\ }in\ \href@noop {} {\emph {\bibinfo {booktitle}
  {2007 Quantum Electronics and Laser Science Conference}}}\ (\bibinfo {year}
  {2007})\ pp.\ \bibinfo {pages} {1--2}\BibitemShut {NoStop}%
\bibitem [{\citenamefont {Ding}\ \emph {et~al.}(2010)\citenamefont {Ding},
  \citenamefont {Singh}, \citenamefont {Plumhof}, \citenamefont {Zander},
  \citenamefont {K\v{r}\'{a}pek}, \citenamefont {Chen}, \citenamefont
  {Benyoucef}, \citenamefont {Zwiller}, \citenamefont {D\"{o}rr}, \citenamefont
  {Bester}, \citenamefont {Rastelli},\ and\ \citenamefont
  {Schmidt}}]{Ding2010}%
  \BibitemOpen
  \bibfield  {author} {\bibinfo {author} {\bibfnamefont {F.}~\bibnamefont
  {Ding}}, \bibinfo {author} {\bibfnamefont {R.}~\bibnamefont {Singh}},
  \bibinfo {author} {\bibfnamefont {J.~D.}\ \bibnamefont {Plumhof}}, \bibinfo
  {author} {\bibfnamefont {T.}~\bibnamefont {Zander}}, \bibinfo {author}
  {\bibfnamefont {V.}~\bibnamefont {K\v{r}\'{a}pek}}, \bibinfo {author}
  {\bibfnamefont {Y.~H.}\ \bibnamefont {Chen}}, \bibinfo {author}
  {\bibfnamefont {M.}~\bibnamefont {Benyoucef}}, \bibinfo {author}
  {\bibfnamefont {V.}~\bibnamefont {Zwiller}}, \bibinfo {author} {\bibfnamefont
  {K.}~\bibnamefont {D\"{o}rr}}, \bibinfo {author} {\bibfnamefont
  {G.}~\bibnamefont {Bester}}, \bibinfo {author} {\bibfnamefont
  {A.}~\bibnamefont {Rastelli}}, \ and\ \bibinfo {author} {\bibfnamefont
  {O.~G.}\ \bibnamefont {Schmidt}},\ }\href {\doibase
  10.1103/PhysRevLett.104.067405} {\bibfield  {journal} {\bibinfo  {journal}
  {Phys. Rev. Lett.}\ }\textbf {\bibinfo {volume} {104}},\ \bibinfo {pages}
  {067405} (\bibinfo {year} {2010})}\BibitemShut {NoStop}%
\bibitem [{\citenamefont {Trotta}\ \emph {et~al.}(2012)\citenamefont {Trotta},
  \citenamefont {Atkinson}, \citenamefont {Plumhof}, \citenamefont {Zallo},
  \citenamefont {Rezaev}, \citenamefont {Kumar}, \citenamefont {Baunack},
  \citenamefont {Schröter}, \citenamefont {Rastelli},\ and\ \citenamefont
  {Schmidt}}]{Trotta2012}%
  \BibitemOpen
  \bibfield  {author} {\bibinfo {author} {\bibfnamefont {R.}~\bibnamefont
  {Trotta}}, \bibinfo {author} {\bibfnamefont {P.}~\bibnamefont {Atkinson}},
  \bibinfo {author} {\bibfnamefont {J.~D.}\ \bibnamefont {Plumhof}}, \bibinfo
  {author} {\bibfnamefont {E.}~\bibnamefont {Zallo}}, \bibinfo {author}
  {\bibfnamefont {R.~O.}\ \bibnamefont {Rezaev}}, \bibinfo {author}
  {\bibfnamefont {S.}~\bibnamefont {Kumar}}, \bibinfo {author} {\bibfnamefont
  {S.}~\bibnamefont {Baunack}}, \bibinfo {author} {\bibfnamefont {J.~R.}\
  \bibnamefont {Schröter}}, \bibinfo {author} {\bibfnamefont {A.}~\bibnamefont
  {Rastelli}}, \ and\ \bibinfo {author} {\bibfnamefont {O.~G.}\ \bibnamefont
  {Schmidt}},\ }\href {\doibase 10.1002/adma.201200537} {\bibfield  {journal}
  {\bibinfo  {journal} {Adv. Mater.}\ }\textbf {\bibinfo {volume} {24}},\
  \bibinfo {pages} {2668} (\bibinfo {year} {2012})}\BibitemShut {NoStop}%
\bibitem [{\citenamefont {Trotta}\ \emph {et~al.}(2013)\citenamefont {Trotta},
  \citenamefont {Zallo}, \citenamefont {Magerl}, \citenamefont {Schmidt},\ and\
  \citenamefont {Rastelli}}]{Trotta2013}%
  \BibitemOpen
  \bibfield  {author} {\bibinfo {author} {\bibfnamefont {R.}~\bibnamefont
  {Trotta}}, \bibinfo {author} {\bibfnamefont {E.}~\bibnamefont {Zallo}},
  \bibinfo {author} {\bibfnamefont {E.}~\bibnamefont {Magerl}}, \bibinfo
  {author} {\bibfnamefont {O.~G.}\ \bibnamefont {Schmidt}}, \ and\ \bibinfo
  {author} {\bibfnamefont {A.}~\bibnamefont {Rastelli}},\ }\href {\doibase
  10.1103/PhysRevB.88.155312} {\bibfield  {journal} {\bibinfo  {journal} {Phys.
  Rev. B}\ }\textbf {\bibinfo {volume} {88}},\ \bibinfo {pages} {155312}
  (\bibinfo {year} {2013})}\BibitemShut {NoStop}%
\bibitem [{\citenamefont {Barth}\ \emph
  {et~al.}(2016{\natexlab{b}})\citenamefont {Barth}, \citenamefont {L\"uker},
  \citenamefont {Vagov}, \citenamefont {Reiter}, \citenamefont {Kuhn},\ and\
  \citenamefont {Axt}}]{Barth2016b}%
  \BibitemOpen
  \bibfield  {author} {\bibinfo {author} {\bibfnamefont {A.~M.}\ \bibnamefont
  {Barth}}, \bibinfo {author} {\bibfnamefont {S.}~\bibnamefont {L\"uker}},
  \bibinfo {author} {\bibfnamefont {A.}~\bibnamefont {Vagov}}, \bibinfo
  {author} {\bibfnamefont {D.~E.}\ \bibnamefont {Reiter}}, \bibinfo {author}
  {\bibfnamefont {T.}~\bibnamefont {Kuhn}}, \ and\ \bibinfo {author}
  {\bibfnamefont {V.~M.}\ \bibnamefont {Axt}},\ }\href {\doibase
  10.1103/PhysRevB.94.045306} {\bibfield  {journal} {\bibinfo  {journal} {Phys.
  Rev. B}\ }\textbf {\bibinfo {volume} {94}},\ \bibinfo {pages} {045306}
  (\bibinfo {year} {2016}{\natexlab{b}})}\BibitemShut {NoStop}%
\bibitem [{\citenamefont {Gl\"assl}\ \emph {et~al.}(2013)\citenamefont
  {Gl\"assl}, \citenamefont {Barth},\ and\ \citenamefont
  {Axt}}]{Glaessl2013Pro}%
  \BibitemOpen
  \bibfield  {author} {\bibinfo {author} {\bibfnamefont {M.}~\bibnamefont
  {Gl\"assl}}, \bibinfo {author} {\bibfnamefont {A.~M.}\ \bibnamefont {Barth}},
  \ and\ \bibinfo {author} {\bibfnamefont {V.~M.}\ \bibnamefont {Axt}},\
  }\href@noop {} {\bibfield  {journal} {\bibinfo  {journal} {Phys.\ Rev.\
  Lett.}\ }\textbf {\bibinfo {volume} {110}},\ \bibinfo {pages} {147401}
  (\bibinfo {year} {2013})}\BibitemShut {NoStop}%
\bibitem [{\citenamefont {Reiter}\ \emph {et~al.}(2014)\citenamefont {Reiter},
  \citenamefont {Kuhn}, \citenamefont {Gl{\"{a}}ssl},\ and\ \citenamefont
  {Axt}}]{Reiter2014}%
  \BibitemOpen
  \bibfield  {author} {\bibinfo {author} {\bibfnamefont {D.~E.}\ \bibnamefont
  {Reiter}}, \bibinfo {author} {\bibfnamefont {T.}~\bibnamefont {Kuhn}},
  \bibinfo {author} {\bibfnamefont {M.}~\bibnamefont {Gl{\"{a}}ssl}}, \ and\
  \bibinfo {author} {\bibfnamefont {V.~M.}\ \bibnamefont {Axt}},\ }\href
  {http://stacks.iop.org/0953-8984/26/i=42/a=423203} {\bibfield  {journal}
  {\bibinfo  {journal} {J. Phys.: Condens. Matter}\ }\textbf {\bibinfo {volume}
  {26}},\ \bibinfo {pages} {423203} (\bibinfo {year} {2014})}\BibitemShut
  {NoStop}%
\bibitem [{\citenamefont {Ardelt}\ \emph {et~al.}(2014)\citenamefont {Ardelt},
  \citenamefont {Hanschke}, \citenamefont {Fischer}, \citenamefont {M\"uller},
  \citenamefont {Kleinkauf}, \citenamefont {Koller}, \citenamefont {Bechtold},
  \citenamefont {Simmet}, \citenamefont {Wierzbowski}, \citenamefont {Riedl},
  \citenamefont {Abstreiter},\ and\ \citenamefont {Finley}}]{Ardelt2014}%
  \BibitemOpen
  \bibfield  {author} {\bibinfo {author} {\bibfnamefont {P.-L.}\ \bibnamefont
  {Ardelt}}, \bibinfo {author} {\bibfnamefont {L.}~\bibnamefont {Hanschke}},
  \bibinfo {author} {\bibfnamefont {K.~A.}\ \bibnamefont {Fischer}}, \bibinfo
  {author} {\bibfnamefont {K.}~\bibnamefont {M\"uller}}, \bibinfo {author}
  {\bibfnamefont {A.}~\bibnamefont {Kleinkauf}}, \bibinfo {author}
  {\bibfnamefont {M.}~\bibnamefont {Koller}}, \bibinfo {author} {\bibfnamefont
  {A.}~\bibnamefont {Bechtold}}, \bibinfo {author} {\bibfnamefont
  {T.}~\bibnamefont {Simmet}}, \bibinfo {author} {\bibfnamefont
  {J.}~\bibnamefont {Wierzbowski}}, \bibinfo {author} {\bibfnamefont
  {H.}~\bibnamefont {Riedl}}, \bibinfo {author} {\bibfnamefont
  {G.}~\bibnamefont {Abstreiter}}, \ and\ \bibinfo {author} {\bibfnamefont
  {J.~J.}\ \bibnamefont {Finley}},\ }\href {\doibase
  10.1103/PhysRevB.90.241404} {\bibfield  {journal} {\bibinfo  {journal} {Phys.
  Rev. B}\ }\textbf {\bibinfo {volume} {90}},\ \bibinfo {pages} {241404(R)}
  (\bibinfo {year} {2014})}\BibitemShut {NoStop}%
\bibitem [{\citenamefont {Bounouar}\ \emph {et~al.}(2015)\citenamefont
  {Bounouar}, \citenamefont {M\"uller}, \citenamefont {Barth}, \citenamefont
  {Gl\"assl}, \citenamefont {Axt},\ and\ \citenamefont
  {Michler}}]{Bounouar2015}%
  \BibitemOpen
  \bibfield  {author} {\bibinfo {author} {\bibfnamefont {S.}~\bibnamefont
  {Bounouar}}, \bibinfo {author} {\bibfnamefont {M.}~\bibnamefont {M\"uller}},
  \bibinfo {author} {\bibfnamefont {A.~M.}\ \bibnamefont {Barth}}, \bibinfo
  {author} {\bibfnamefont {M.}~\bibnamefont {Gl\"assl}}, \bibinfo {author}
  {\bibfnamefont {V.~M.}\ \bibnamefont {Axt}}, \ and\ \bibinfo {author}
  {\bibfnamefont {P.}~\bibnamefont {Michler}},\ }\href {\doibase
  10.1103/PhysRevB.91.161302} {\bibfield  {journal} {\bibinfo  {journal} {Phys.
  Rev. B}\ }\textbf {\bibinfo {volume} {91}},\ \bibinfo {pages} {161302(R)}
  (\bibinfo {year} {2015})}\BibitemShut {NoStop}%
\bibitem [{\citenamefont {Quilter}\ \emph {et~al.}(2015)\citenamefont
  {Quilter}, \citenamefont {Brash}, \citenamefont {Liu}, \citenamefont
  {Gl\"assl}, \citenamefont {Barth}, \citenamefont {Axt}, \citenamefont
  {Ramsay}, \citenamefont {Skolnick},\ and\ \citenamefont {Fox}}]{Quilter2015}%
  \BibitemOpen
  \bibfield  {author} {\bibinfo {author} {\bibfnamefont {J.~H.}\ \bibnamefont
  {Quilter}}, \bibinfo {author} {\bibfnamefont {A.~J.}\ \bibnamefont {Brash}},
  \bibinfo {author} {\bibfnamefont {F.}~\bibnamefont {Liu}}, \bibinfo {author}
  {\bibfnamefont {M.}~\bibnamefont {Gl\"assl}}, \bibinfo {author}
  {\bibfnamefont {A.~M.}\ \bibnamefont {Barth}}, \bibinfo {author}
  {\bibfnamefont {V.~M.}\ \bibnamefont {Axt}}, \bibinfo {author} {\bibfnamefont
  {A.~J.}\ \bibnamefont {Ramsay}}, \bibinfo {author} {\bibfnamefont {M.~S.}\
  \bibnamefont {Skolnick}}, \ and\ \bibinfo {author} {\bibfnamefont {A.~M.}\
  \bibnamefont {Fox}},\ }\href {\doibase 10.1103/PhysRevLett.114.137401}
  {\bibfield  {journal} {\bibinfo  {journal} {Phys. Rev. Lett.}\ }\textbf
  {\bibinfo {volume} {114}},\ \bibinfo {pages} {137401} (\bibinfo {year}
  {2015})}\BibitemShut {NoStop}%
\bibitem [{\citenamefont {Reindl}\ \emph {et~al.}(2017)\citenamefont {Reindl},
  \citenamefont {J\"ons}, \citenamefont {Huber}, \citenamefont {Schimpf},
  \citenamefont {Huo}, \citenamefont {Zwiller}, \citenamefont {Rastelli},\ and\
  \citenamefont {Trotta}}]{Reindl2017}%
  \BibitemOpen
  \bibfield  {author} {\bibinfo {author} {\bibfnamefont {M.}~\bibnamefont
  {Reindl}}, \bibinfo {author} {\bibfnamefont {K.~D.}\ \bibnamefont {J\"ons}},
  \bibinfo {author} {\bibfnamefont {D.}~\bibnamefont {Huber}}, \bibinfo
  {author} {\bibfnamefont {C.}~\bibnamefont {Schimpf}}, \bibinfo {author}
  {\bibfnamefont {Y.}~\bibnamefont {Huo}}, \bibinfo {author} {\bibfnamefont
  {V.}~\bibnamefont {Zwiller}}, \bibinfo {author} {\bibfnamefont
  {A.}~\bibnamefont {Rastelli}}, \ and\ \bibinfo {author} {\bibfnamefont
  {R.}~\bibnamefont {Trotta}},\ }\href {\doibase 10.1021/acs.nanolett.7b00777}
  {\bibfield  {journal} {\bibinfo  {journal} {Nano Lett.}\ }\textbf {\bibinfo
  {volume} {17}},\ \bibinfo {pages} {4090} (\bibinfo {year}
  {2017})}\BibitemShut {NoStop}%
\bibitem [{\citenamefont {Ates}\ \emph {et~al.}(2009)\citenamefont {Ates},
  \citenamefont {Ulrich}, \citenamefont {Ulhaq}, \citenamefont {Reitzenstein},
  \citenamefont {L{\"o}ffler}, \citenamefont {H{\"o}fling}, \citenamefont
  {Forchel},\ and\ \citenamefont {Michler}}]{Ates2009}%
  \BibitemOpen
  \bibfield  {author} {\bibinfo {author} {\bibfnamefont {S.}~\bibnamefont
  {Ates}}, \bibinfo {author} {\bibfnamefont {S.~M.}\ \bibnamefont {Ulrich}},
  \bibinfo {author} {\bibfnamefont {A.}~\bibnamefont {Ulhaq}}, \bibinfo
  {author} {\bibfnamefont {S.}~\bibnamefont {Reitzenstein}}, \bibinfo {author}
  {\bibfnamefont {A.}~\bibnamefont {L{\"o}ffler}}, \bibinfo {author}
  {\bibfnamefont {S.}~\bibnamefont {H{\"o}fling}}, \bibinfo {author}
  {\bibfnamefont {A.}~\bibnamefont {Forchel}}, \ and\ \bibinfo {author}
  {\bibfnamefont {P.}~\bibnamefont {Michler}},\ }\href {\doibase
  10.1038/nphoton.2009.215} {\bibfield  {journal} {\bibinfo  {journal} {Nature
  Photonics}\ }\textbf {\bibinfo {volume} {3}},\ \bibinfo {pages} {724}
  (\bibinfo {year} {2009})}\BibitemShut {NoStop}%
\bibitem [{\citenamefont {Naesby}\ \emph {et~al.}(2008)\citenamefont {Naesby},
  \citenamefont {Suhr}, \citenamefont {Kristensen},\ and\ \citenamefont
  {M\o{}rk}}]{Naesby2008}%
  \BibitemOpen
  \bibfield  {author} {\bibinfo {author} {\bibfnamefont {A.}~\bibnamefont
  {Naesby}}, \bibinfo {author} {\bibfnamefont {T.}~\bibnamefont {Suhr}},
  \bibinfo {author} {\bibfnamefont {P.~T.}\ \bibnamefont {Kristensen}}, \ and\
  \bibinfo {author} {\bibfnamefont {J.}~\bibnamefont {M\o{}rk}},\ }\href
  {\doibase 10.1103/PhysRevA.78.045802} {\bibfield  {journal} {\bibinfo
  {journal} {Phys. Rev. A}\ }\textbf {\bibinfo {volume} {78}},\ \bibinfo
  {pages} {045802} (\bibinfo {year} {2008})}\BibitemShut {NoStop}%
\bibitem [{\citenamefont {Hohenester}\ \emph {et~al.}(2009)\citenamefont
  {Hohenester}, \citenamefont {Laucht}, \citenamefont {Kaniber}, \citenamefont
  {Hauke}, \citenamefont {Neumann}, \citenamefont {Mohtashami}, \citenamefont
  {Seliger}, \citenamefont {Bichler},\ and\ \citenamefont
  {Finley}}]{Hohenester2009}%
  \BibitemOpen
  \bibfield  {author} {\bibinfo {author} {\bibfnamefont {U.}~\bibnamefont
  {Hohenester}}, \bibinfo {author} {\bibfnamefont {A.}~\bibnamefont {Laucht}},
  \bibinfo {author} {\bibfnamefont {M.}~\bibnamefont {Kaniber}}, \bibinfo
  {author} {\bibfnamefont {N.}~\bibnamefont {Hauke}}, \bibinfo {author}
  {\bibfnamefont {A.}~\bibnamefont {Neumann}}, \bibinfo {author} {\bibfnamefont
  {A.}~\bibnamefont {Mohtashami}}, \bibinfo {author} {\bibfnamefont
  {M.}~\bibnamefont {Seliger}}, \bibinfo {author} {\bibfnamefont
  {M.}~\bibnamefont {Bichler}}, \ and\ \bibinfo {author} {\bibfnamefont
  {J.~J.}\ \bibnamefont {Finley}},\ }\href {\doibase
  10.1103/PhysRevB.80.201311} {\bibfield  {journal} {\bibinfo  {journal} {Phys.
  Rev. B}\ }\textbf {\bibinfo {volume} {80}},\ \bibinfo {pages} {201311(R)}
  (\bibinfo {year} {2009})}\BibitemShut {NoStop}%
\bibitem [{\citenamefont {Hohenester}(2010)}]{Hohenester2010}%
  \BibitemOpen
  \bibfield  {author} {\bibinfo {author} {\bibfnamefont {U.}~\bibnamefont
  {Hohenester}},\ }\href {\doibase 10.1103/PhysRevB.81.155303} {\bibfield
  {journal} {\bibinfo  {journal} {Phys. Rev. B}\ }\textbf {\bibinfo {volume}
  {81}},\ \bibinfo {pages} {155303} (\bibinfo {year} {2010})}\BibitemShut
  {NoStop}%
\bibitem [{\citenamefont {Majumdar}\ \emph {et~al.}(2011)\citenamefont
  {Majumdar}, \citenamefont {Kim}, \citenamefont {Gong}, \citenamefont
  {Bajcsy},\ and\ \citenamefont {Vu\ifmmode \check{c}\else
  \v{c}\fi{}kovi\ifmmode~\acute{c}\else \'{c}\fi{}}}]{Majumdar2011}%
  \BibitemOpen
  \bibfield  {author} {\bibinfo {author} {\bibfnamefont {A.}~\bibnamefont
  {Majumdar}}, \bibinfo {author} {\bibfnamefont {E.~D.}\ \bibnamefont {Kim}},
  \bibinfo {author} {\bibfnamefont {Y.}~\bibnamefont {Gong}}, \bibinfo {author}
  {\bibfnamefont {M.}~\bibnamefont {Bajcsy}}, \ and\ \bibinfo {author}
  {\bibfnamefont {J.}~\bibnamefont {Vu\ifmmode \check{c}\else
  \v{c}\fi{}kovi\ifmmode~\acute{c}\else \'{c}\fi{}}},\ }\href {\doibase
  10.1103/PhysRevB.84.085309} {\bibfield  {journal} {\bibinfo  {journal} {Phys.
  Rev. B}\ }\textbf {\bibinfo {volume} {84}},\ \bibinfo {pages} {085309}
  (\bibinfo {year} {2011})}\BibitemShut {NoStop}%
\bibitem [{\citenamefont {Hughes}\ \emph {et~al.}(2011)\citenamefont {Hughes},
  \citenamefont {Yao}, \citenamefont {Milde}, \citenamefont {Knorr},
  \citenamefont {Dalacu}, \citenamefont {Mnaymneh}, \citenamefont {Sazonova},
  \citenamefont {Poole}, \citenamefont {Aers}, \citenamefont {Lapointe},
  \citenamefont {Cheriton},\ and\ \citenamefont {Williams}}]{Hughes2011}%
  \BibitemOpen
  \bibfield  {author} {\bibinfo {author} {\bibfnamefont {S.}~\bibnamefont
  {Hughes}}, \bibinfo {author} {\bibfnamefont {P.}~\bibnamefont {Yao}},
  \bibinfo {author} {\bibfnamefont {F.}~\bibnamefont {Milde}}, \bibinfo
  {author} {\bibfnamefont {A.}~\bibnamefont {Knorr}}, \bibinfo {author}
  {\bibfnamefont {D.}~\bibnamefont {Dalacu}}, \bibinfo {author} {\bibfnamefont
  {K.}~\bibnamefont {Mnaymneh}}, \bibinfo {author} {\bibfnamefont
  {V.}~\bibnamefont {Sazonova}}, \bibinfo {author} {\bibfnamefont {P.~J.}\
  \bibnamefont {Poole}}, \bibinfo {author} {\bibfnamefont {G.~C.}\ \bibnamefont
  {Aers}}, \bibinfo {author} {\bibfnamefont {J.}~\bibnamefont {Lapointe}},
  \bibinfo {author} {\bibfnamefont {R.}~\bibnamefont {Cheriton}}, \ and\
  \bibinfo {author} {\bibfnamefont {R.~L.}\ \bibnamefont {Williams}},\ }\href
  {\doibase 10.1103/PhysRevB.83.165313} {\bibfield  {journal} {\bibinfo
  {journal} {Phys. Rev. B}\ }\textbf {\bibinfo {volume} {83}},\ \bibinfo
  {pages} {165313} (\bibinfo {year} {2011})}\BibitemShut {NoStop}%
\bibitem [{\citenamefont {Florian}\ \emph {et~al.}(2013)\citenamefont
  {Florian}, \citenamefont {Gartner}, \citenamefont {Gies},\ and\ \citenamefont
  {Jahnke}}]{Florian2013}%
  \BibitemOpen
  \bibfield  {author} {\bibinfo {author} {\bibfnamefont {M.}~\bibnamefont
  {Florian}}, \bibinfo {author} {\bibfnamefont {P.}~\bibnamefont {Gartner}},
  \bibinfo {author} {\bibfnamefont {C.}~\bibnamefont {Gies}}, \ and\ \bibinfo
  {author} {\bibfnamefont {F.}~\bibnamefont {Jahnke}},\ }\href {\doibase
  10.1088/1367-2630/15/3/035019} {\bibfield  {journal} {\bibinfo  {journal}
  {New Journal of Physics}\ }\textbf {\bibinfo {volume} {15}},\ \bibinfo
  {pages} {035019} (\bibinfo {year} {2013})}\BibitemShut {NoStop}%
\bibitem [{\citenamefont {Calic}\ \emph {et~al.}(2011)\citenamefont {Calic},
  \citenamefont {Gallo}, \citenamefont {Felici}, \citenamefont {Atlasov},
  \citenamefont {Dwir}, \citenamefont {Rudra}, \citenamefont {Biasiol},
  \citenamefont {Sorba}, \citenamefont {Tarel}, \citenamefont {Savona},\ and\
  \citenamefont {Kapon}}]{Calic2011}%
  \BibitemOpen
  \bibfield  {author} {\bibinfo {author} {\bibfnamefont {M.}~\bibnamefont
  {Calic}}, \bibinfo {author} {\bibfnamefont {P.}~\bibnamefont {Gallo}},
  \bibinfo {author} {\bibfnamefont {M.}~\bibnamefont {Felici}}, \bibinfo
  {author} {\bibfnamefont {K.~A.}\ \bibnamefont {Atlasov}}, \bibinfo {author}
  {\bibfnamefont {B.}~\bibnamefont {Dwir}}, \bibinfo {author} {\bibfnamefont
  {A.}~\bibnamefont {Rudra}}, \bibinfo {author} {\bibfnamefont
  {G.}~\bibnamefont {Biasiol}}, \bibinfo {author} {\bibfnamefont
  {L.}~\bibnamefont {Sorba}}, \bibinfo {author} {\bibfnamefont
  {G.}~\bibnamefont {Tarel}}, \bibinfo {author} {\bibfnamefont
  {V.}~\bibnamefont {Savona}}, \ and\ \bibinfo {author} {\bibfnamefont
  {E.}~\bibnamefont {Kapon}},\ }\href {\doibase 10.1103/PhysRevLett.106.227402}
  {\bibfield  {journal} {\bibinfo  {journal} {Phys. Rev. Lett.}\ }\textbf
  {\bibinfo {volume} {106}},\ \bibinfo {pages} {227402} (\bibinfo {year}
  {2011})}\BibitemShut {NoStop}%
\bibitem [{\citenamefont {Laucht}\ \emph {et~al.}(2011)\citenamefont {Laucht},
  \citenamefont {Hauke}, \citenamefont {Neumann}, \citenamefont {Günthner},
  \citenamefont {Hofbauer}, \citenamefont {Mohtashami}, \citenamefont
  {Müller}, \citenamefont {Böhm}, \citenamefont {Bichler}, \citenamefont
  {Amann}, \citenamefont {Kaniber},\ and\ \citenamefont {Finley}}]{Laucht2011}%
  \BibitemOpen
  \bibfield  {author} {\bibinfo {author} {\bibfnamefont {A.}~\bibnamefont
  {Laucht}}, \bibinfo {author} {\bibfnamefont {N.}~\bibnamefont {Hauke}},
  \bibinfo {author} {\bibfnamefont {A.}~\bibnamefont {Neumann}}, \bibinfo
  {author} {\bibfnamefont {T.}~\bibnamefont {Günthner}}, \bibinfo {author}
  {\bibfnamefont {F.}~\bibnamefont {Hofbauer}}, \bibinfo {author}
  {\bibfnamefont {A.}~\bibnamefont {Mohtashami}}, \bibinfo {author}
  {\bibfnamefont {K.}~\bibnamefont {Müller}}, \bibinfo {author} {\bibfnamefont
  {G.}~\bibnamefont {Böhm}}, \bibinfo {author} {\bibfnamefont
  {M.}~\bibnamefont {Bichler}}, \bibinfo {author} {\bibfnamefont {M.-C.}\
  \bibnamefont {Amann}}, \bibinfo {author} {\bibfnamefont {M.}~\bibnamefont
  {Kaniber}}, \ and\ \bibinfo {author} {\bibfnamefont {J.~J.}\ \bibnamefont
  {Finley}},\ }\href {\doibase 10.1063/1.3576137} {\bibfield  {journal}
  {\bibinfo  {journal} {Journal of Applied Physics}\ }\textbf {\bibinfo
  {volume} {109}},\ \bibinfo {pages} {102404} (\bibinfo {year}
  {2011})}\BibitemShut {NoStop}%
\bibitem [{\citenamefont {Makri}\ and\ \citenamefont
  {Makarov}(1995{\natexlab{a}})}]{Makri1995a}%
  \BibitemOpen
  \bibfield  {author} {\bibinfo {author} {\bibfnamefont {N.}~\bibnamefont
  {Makri}}\ and\ \bibinfo {author} {\bibfnamefont {D.~E.}\ \bibnamefont
  {Makarov}},\ }\href {\doibase 10.1063/1.469508} {\bibfield  {journal}
  {\bibinfo  {journal} {J. Chem. Phys.}\ }\textbf {\bibinfo {volume} {102}},\
  \bibinfo {pages} {4600} (\bibinfo {year} {1995}{\natexlab{a}})}\BibitemShut
  {NoStop}%
\bibitem [{\citenamefont {Makri}\ and\ \citenamefont
  {Makarov}(1995{\natexlab{b}})}]{Makri1995b}%
  \BibitemOpen
  \bibfield  {author} {\bibinfo {author} {\bibfnamefont {N.}~\bibnamefont
  {Makri}}\ and\ \bibinfo {author} {\bibfnamefont {D.~E.}\ \bibnamefont
  {Makarov}},\ }\href {\doibase 10.1063/1.469509} {\bibfield  {journal}
  {\bibinfo  {journal} {J. Chem. Phys.}\ }\textbf {\bibinfo {volume} {102}},\
  \bibinfo {pages} {4611} (\bibinfo {year} {1995}{\natexlab{b}})}\BibitemShut
  {NoStop}%
\bibitem [{\citenamefont {Krummheuer}\ \emph {et~al.}(2005)\citenamefont
  {Krummheuer}, \citenamefont {Axt}, \citenamefont {Kuhn}, \citenamefont
  {D'Amico},\ and\ \citenamefont {Rossi}}]{Krummheuer2005}%
  \BibitemOpen
  \bibfield  {author} {\bibinfo {author} {\bibfnamefont {B.}~\bibnamefont
  {Krummheuer}}, \bibinfo {author} {\bibfnamefont {V.~M.}\ \bibnamefont {Axt}},
  \bibinfo {author} {\bibfnamefont {T.}~\bibnamefont {Kuhn}}, \bibinfo {author}
  {\bibfnamefont {I.}~\bibnamefont {D'Amico}}, \ and\ \bibinfo {author}
  {\bibfnamefont {F.}~\bibnamefont {Rossi}},\ }\href {\doibase
  10.1103/PhysRevB.71.235329} {\bibfield  {journal} {\bibinfo  {journal} {Phys.
  Rev. B}\ }\textbf {\bibinfo {volume} {71}},\ \bibinfo {pages} {235329}
  (\bibinfo {year} {2005})}\BibitemShut {NoStop}%
\bibitem [{\citenamefont {Schneider}\ \emph {et~al.}(2016)\citenamefont
  {Schneider}, \citenamefont {Gold}, \citenamefont {Reitzenstein},
  \citenamefont {H{\"o}fling},\ and\ \citenamefont {Kamp}}]{Schneider2016}%
  \BibitemOpen
  \bibfield  {author} {\bibinfo {author} {\bibfnamefont {C.}~\bibnamefont
  {Schneider}}, \bibinfo {author} {\bibfnamefont {P.}~\bibnamefont {Gold}},
  \bibinfo {author} {\bibfnamefont {S.}~\bibnamefont {Reitzenstein}}, \bibinfo
  {author} {\bibfnamefont {S.}~\bibnamefont {H{\"o}fling}}, \ and\ \bibinfo
  {author} {\bibfnamefont {M.}~\bibnamefont {Kamp}},\ }\href {\doibase
  10.1007/s00340-015-6283-x} {\bibfield  {journal} {\bibinfo  {journal}
  {Applied Physics B}\ }\textbf {\bibinfo {volume} {122}},\ \bibinfo {pages}
  {19} (\bibinfo {year} {2016})}\BibitemShut {NoStop}%
\end{thebibliography}%
\end{document}